\documentclass[acmtog]{acmart}
\acmSubmissionID{0105}
\setcitestyle{square}

\usepackage[ruled]{algorithm2e} 
\usepackage{booktabs} 
\usepackage{graphicx}
\usepackage{makecell}
\usepackage{multirow}
\usepackage{wrapfig}

\newcommand{\Eq}[1]   {Equation \ref{eq:#1}}
\newcommand{\Fig}[1]  {Figure \ref{fig:#1}}
\newcommand{\Tbl}[1]  {Table \ref{tbl:#1}}
\newcommand{\Sec}[1]  {Section \ref{sec:#1}}

\setcopyright{none}

\acmDOI{XXXXXXXXXX}

\acmISBN{XXXXXXXXXXXXX}

\newcommand{\rev}[1]{{#1}}

\title{ComplementMe:\\Weakly-Supervised Component Suggestions for 3D Modeling}

\author{Minhyuk Sung}
\email{mhsung@cs.stanford.edu}
\affiliation{%
  \institution{Stanford University}
  \department{Computer Science}
  \city{Stanford}
  \state{CA}
  \postcode{94305}
  \country{USA}
}

\author{Hao Su}
\affiliation{%
  \institution{Stanford University and University of California San Diego}
}

\author{Vladimir G. Kim}
\affiliation{%
	\institution{Adobe Research}
}

\author{Siddhartha Chaudhuri}
\affiliation{%
	\institution{IIT Bombay}
}

\author{Leonidas Guibas}
\email{guibas@cs.stanford.edu}
\affiliation{%
  \institution{Stanford University}
  \department{Computer Science}
  \city{Stanford}
  \state{CA}
  \postcode{94305}
  \country{USA}
}

\renewcommand{\shortauthors}{}

\begin{document}


\begin{abstract}
\rev{Assembly-based tools provide a powerful modeling paradigm for non-expert shape designers. However, choosing a component from a large shape repository and aligning it to a partial assembly can become a daunting task. 
In this paper we describe novel neural network architectures for suggesting complementary components and their placement for an incomplete 3D part assembly. Unlike most existing techniques, our networks are trained on unlabeled data obtained from public online repositories, and do not rely on consistent part segmentations or labels. 
Absence of labels poses a challenge in indexing the database of parts for the retrieval. We address it by jointly training embedding and retrieval networks, where the first indexes parts by mapping them to a low-dimensional feature space, and the second maps partial assemblies to appropriate complements. The combinatorial nature of part arrangements poses another challenge, since the retrieval network is not a function: several complements can be appropriate for the same input. Thus, instead of predicting a single output, we train our network to predict a probability distribution over the space of part embeddings. This allows our method to deal with ambiguities and naturally enables a UI that seamlessly integrates user preferences into the design process. 
%
%
We demonstrate that our method can be used to design complex shapes with minimal or no user input. 
To evaluate our approach we develop a novel benchmark for component suggestion systems demonstrating significant improvement over state-of-the-art techniques. 
}
\end{abstract}



\thanks{We thank the anonymous reviewers for their comments and suggestions. This project was supported by NSF grants IIS-1528025 and DMS-1521608, MURI award N00014-13-1-0341, a Google focused research award, the Korea Foundation for Advanced Studies, and gifts from the Adobe systems and Autodesk corporations.}

\begin{teaserfigure}
\centering
\includegraphics[width=\textwidth]{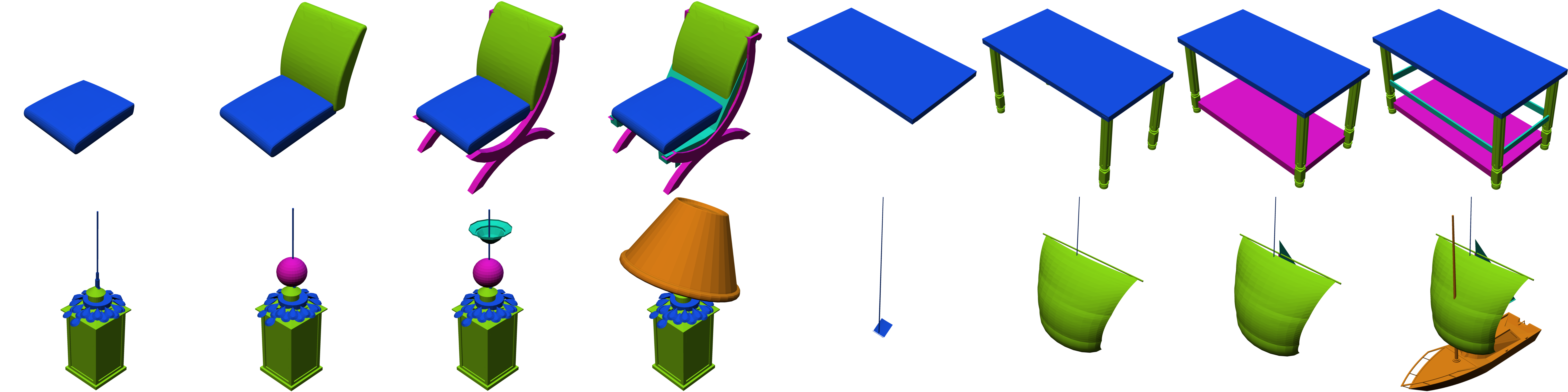}
\caption{Automatic shape synthesis via incremental component assembly. Note that a component and its placement is suggested automatically at every iteration, given only a partial shape from the previous iteration.}
\label{fig:teaser}
\end{teaserfigure}

\maketitle


\section{Introduction}
\label{sec:introduction}


Geometric modeling is essential for populating virtual environments as well as for designing real objects. 
%
%
Yet creating 3D models from scratch is a tedious and time-consuming process that requires substantial expertise. 
To address this challenge, Funkhouser et al.~\shortcite{Funkhouser:2004} put forth the idea of re-using parts of existing 3D models to generate new content. 
To alleviate the burden of finding and segmenting geometric regions, Chaudhuri et al.~\shortcite{Chaudhuri:2011} proposed an interface for part-wise shape assembly, which reduces the user interaction to component selection and placement. Their suggestion model was trained on a heavily supervised dataset, where every shape was segmented into a consistent set of parts with semantic part labels. Even at a very coarse part level, significant expense has to be incurred with the help of crowd-sourced workers and active learning techniques~\cite{yi:2016}.
%


In this work we propose a novel component suggestion approach that does not require explicit part annotations. While our method still requires unlabeled components, this is a much weaker requirement, as it has been observed before that these decompositions can be done automatically~\cite{Chaudhuri:2011}. In this work, we also leverage the observation of Yi et al.~\shortcite{Li:2017} that models that come from online repositories, such as the 3D Warehouse~\cite{Warehouse}, already have some segmentations (based on connected components and scene graph nodes)  that often align with natural part boundaries. Despite the fact that these components are inconsistent and unlabeled we can still train a model for component suggestion, because given some (partial) shape assembly we know exactly which components are missing and where they need to be placed.
%


\rev{We propose novel neural network architectures for suggesting complementary components and their locations given a partially assembled shape. Our networks use unordered point clouds to represent geometry~\cite{Qi:2017}, which makes them widely applicable\footnote{In particular, this makes it possible to easily integrate our approach within other extant design systems.}.  
There are two main challenges in training the retrieval network. First, since we do not require consistent segmentations and labels our network needs to index the parts for retrieval. Thus, we jointly train two networks, an embedding network that indexes the parts by mapping them to a low-dimensional latent space, and a retrieval network that maps a partial assembly to the appropriate subspace of complements. These networks are trained together from triplets of examples: a partial assembly, a correct complement, and an incorrect complement. We use contrastive loss to ensure that correct and incorrect complements are separated by a margin, which favors embeddings that are compatible with the retrieval network predictions. The second challenge is that multiple design options can complement each partial assembly (e.g., one can add either legs, or a back, or arm rests to a seat of a chair). We address this challenge by predicting a probability distribution over the space of all plausible predictions, which we model as a mixture of Gaussians with confidence weights. This enables us to train a network that suggests multiple plausible solutions simultaneously, even though every training pair provides only one solution. Finally, the location prediction network takes a partial assembly and a complementary component and outputs possible placements for that component. 
}

We demonstrate that our method leads to a modeling tool that requires minimal or no user input. We also propose a novel benchmark to evaluate the performance of component suggestion methods. In that setting, our approach outperforms state-of-the-art retrieval techniques that do not rely on heavily curated datasets.



\section{Related Work}
\label{sec:related_work}

We review related work on assembly-based modeling and recent uses of neural networks for geometric modeling.

{\noindent \bfseries  3D modeling by assembly.}
Funkhouser et al.~\shortcite{Funkhouser:2004} pioneered the idea of creating 3D models by assembling parts segmented from shapes in a repository. Subsequent interfaces reduce the amount of tedious manual segmentation by using a heavily curated repository of objects pre-segmented into labeled parts~\cite{Chaudhuri:2011,Kalogerakis:2012}. They proposed a probabilistic graphical model to reason about which parts can complement one another.
%
Part assemblies can also be used to create plausible complete objects from partial point clouds. For example, Shen et al.~\shortcite{Shen:2012} detect and fill missing components by aligning the input to 3D models in the database. Sung et al.~\shortcite{Sung:2015} fit structure templates to the partial scan data to leverage both symmetry and part retrieval for completion.
These part-based models rely on a database of consistently segmented shapes with part labels, which limits the applicability of these techniques as they incur significant data annotation costs~\cite{yi:2016}.

There are two notable exceptions. Jaiswal et al.~\shortcite{Jaiswal:2016} used factor graphs to model pairwise compatibilities when suggesting a new part. Their suggestions are based only on pairwise relationships, rendering this method less suitable for holistic reasoning. Chaudhuri and Koltun~\shortcite{Chaudhuri:2010} proposed a method that retrieves partially similar shapes and detect components that can be added to the existing assembly. They assumed that the coarse shape is mostly complete, so that global shape descriptors can reliably retrieve a structurally similar model, and that part placement will not change significantly from the retrieved model.  While these techniques also do not require part labels and consistent segmentations, unlike our approach, they do not learn how to  predict parts. There are several issues associated with that. First, hand-crafted shape descriptors, parameters, and weights that they use in their systems might have to be adapted as one switches to new dataset. Second, it is challenging for these systems to learn what a complete target shape in a particular category looks like. In contrast, our method uses neural networks to learn an appropriate shape representation to map a partial assembly to complementary parts and their respective positions. It does not require manual parameter tuning and can easily apply to a wide range of shape categories. 

\vspace{0.5cm}
{\noindent \bfseries Neural networks for 3D modeling.}
Several recent techniques use neural networks for modeling 3D shapes.  A direct extension of image synthesis is 3D volume synthesis, where previous work explored synthesizing volumes from depth~\cite{Wu:2015}, images~\cite{Choy:2016,Grant:2016}, or both~\cite{Tulsiani:2017}. Other output 3D representations include skeletons~\cite{Wu:2016c}, graph-based structures~\cite{Kong:2017}, and point clouds~\cite{Fan:2017}.

In this work we demonstrate that neural networks can also be used for incremental interactive shape assembly from parts. Since our geometry representation focuses on retrieving appropriate components from the repository instead of synthesizing geometry from scratch, we are able to create high fidelity 3D meshes.

Since our assembly process relies on training a component retrieval network, our method is also related to learning shape embeddings. Previous techniques learned embeddings for different purposes: \cite{Girdhar:2016} for reconstructing 3D from 2D, \cite{Sharma:2016} for denoising, \cite{Wu:2016} for synthesizing shapes, and \cite{Li:2017} for detecting hierarchical part structures. 
We introduce a different embedding designed specifically for our retrieval problem. Our approach jointly learns to embed complementing components that occur in similar context nearby, and learns to map partial objects to their complements.


\begin{figure*}[t]
\centering
\includegraphics[width=0.8\linewidth]{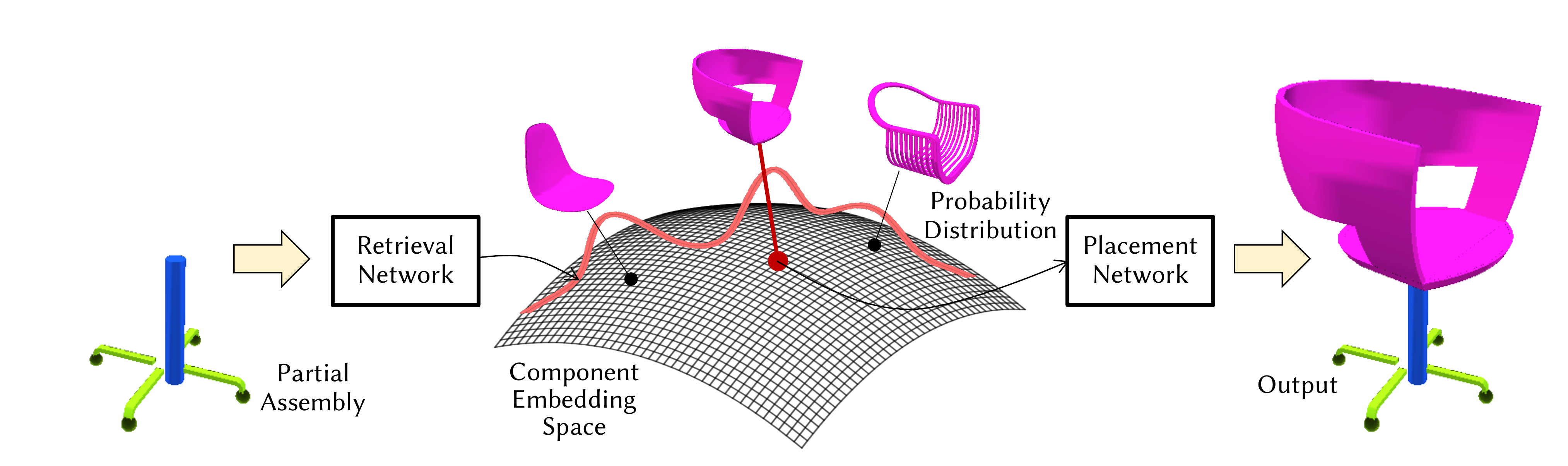}
\caption{\rev{Overview of the retrieval process at test time. From the given partial assembly, the retrieval network predicts a probability distribution (modeled as a mixture of Gaussians) over the component embedding space. Suggested complementary components are sampled from the predicted distribution, and then the placement network predicts their positions with respect to the query assembly. }}
\label{fig:overview}
\end{figure*}

\section{Overview}
\label{sec:overview}
Incremental assembly-based modeling systems require two key technical components: part retrieval and part placement. In this work we provide solutions to both of these problems. In particular, given a partial object, our method proposes a set of potential complementary components from a repository of 3D shapes and a placement of each component.  The goal is to retrieve components that are \emph{compatible} with the partial assembly in style, functionality and other factors; while simultaneously are as \emph{diverse} as possible to leave more options to the designer. We also need to predict positions for these components so that they form a valid shape with the partial assembly. These are challenging problems that require human-level understanding of objects, and thus we propose learning-based approaches for generating these proposals. 

\rev{Our first challenge is to obtain the training data: pairs of geometries including a partial 3D assembly and potential complementing components. We use the 3D models from ShapeNet~\cite{Chang:2015}, a large-scale online repository, to create these pairs. We first need to decompose these objects into components, which form the basic unit of our system. Unlike most previous works, we do not require these decompositions to be consistent across shapes, have explicit correspondences, or have labels. Similar to Chaudhuri et al.~\shortcite{Chaudhuri:2011}, we could use existing segmentation algorithms. However, following the observation of Yi et al.~\shortcite{Yi:2017} we found that most shapes in these repositories are composed of connected components that mostly align with natural part boundaries. Thus, we propose a simple data pre-processing procedure that merges small and repetitive components, and uses the resulting larger parts.  While we could train directly on these parts by picking a subset of components and trying to predict the rest, we found that it is very unintuitive to predict a part that is not attached to the current assembly, and thus use a proximity-based graph of the processed components to avoid training on disconnected examples. We describe this step in more details in Section~\ref{sec:data_preprocessing}.}

We use this data to train a neural network for selecting complementary components, and use point clouds to represent shape geometry~\cite{Qi:2017}. To train our network we pick a random connected subgraph of components as input and use all remaining components adjacent to the subgraph as training examples. For example, given a single chair seat, a back, a leg, or arm-rests are all correct suggestions. Furthermore, in practice any of these parts in the style that is compatible to the seat can be valid retrievals. 
This means that the mapping from our inputs $X$ to the outputs $Y$ is not a function since it has multiple output values, and thus cannot be modeled with a simple regression. 

In this work we address two fundamental challenges associated with the retrieval problem: how to model ambiguity in retrievals and how to index parts.
To address the first challenge we propose a \emph{retrieval network} architecture that produces a conditional probability distribution $P(Y|X)$ modeled as a Gaussian mixture for the output.  Our network is designed based on the Mixture Density Network \cite{Bishop:1994}.  This method enables us to retrieve a diverse set of plausible results, as detailed in Section~\ref{sec:retrieval_network}.
To address the second challenge we learn a low-dimensional embedding of all parts to encode the retrieved result $Y$. Then, proposing new components corresponds to sampling a few coordinates in this embedding space according to $P(Y|X)$. While one could use a fixed embedding space (e.g., based on shape descriptors), we learn this embedding by training an \emph{embedding network} that aims to embed compatible complementary parts that share functional and stylistic features nearby (Section~\ref{sec:embedding_network}).
We use a form of contrastive loss to jointly train the retrieval and embedding networks (Section~\ref{sec:joint_training}). 

Finally, we address the challenge of placing the retrieved part in the context of the partial query by training a regression \emph{placement network} that uses both the partial object and a complementary component as an input and the true position of the component as a training example (Section~\ref{sec:placement_network}).

During the incremental assembly design, we first run our retrieval network to obtain a set of high-probability components, and then run the placement network on each component to generate a gallery of potential assembly candidates placed with respect to the input object (see Figure~\ref{fig:overview}).







\section{Data Preprocessing}
\label{sec:data_preprocessing}

\begin{figure}[b]
\centering
\includegraphics[width=\linewidth]{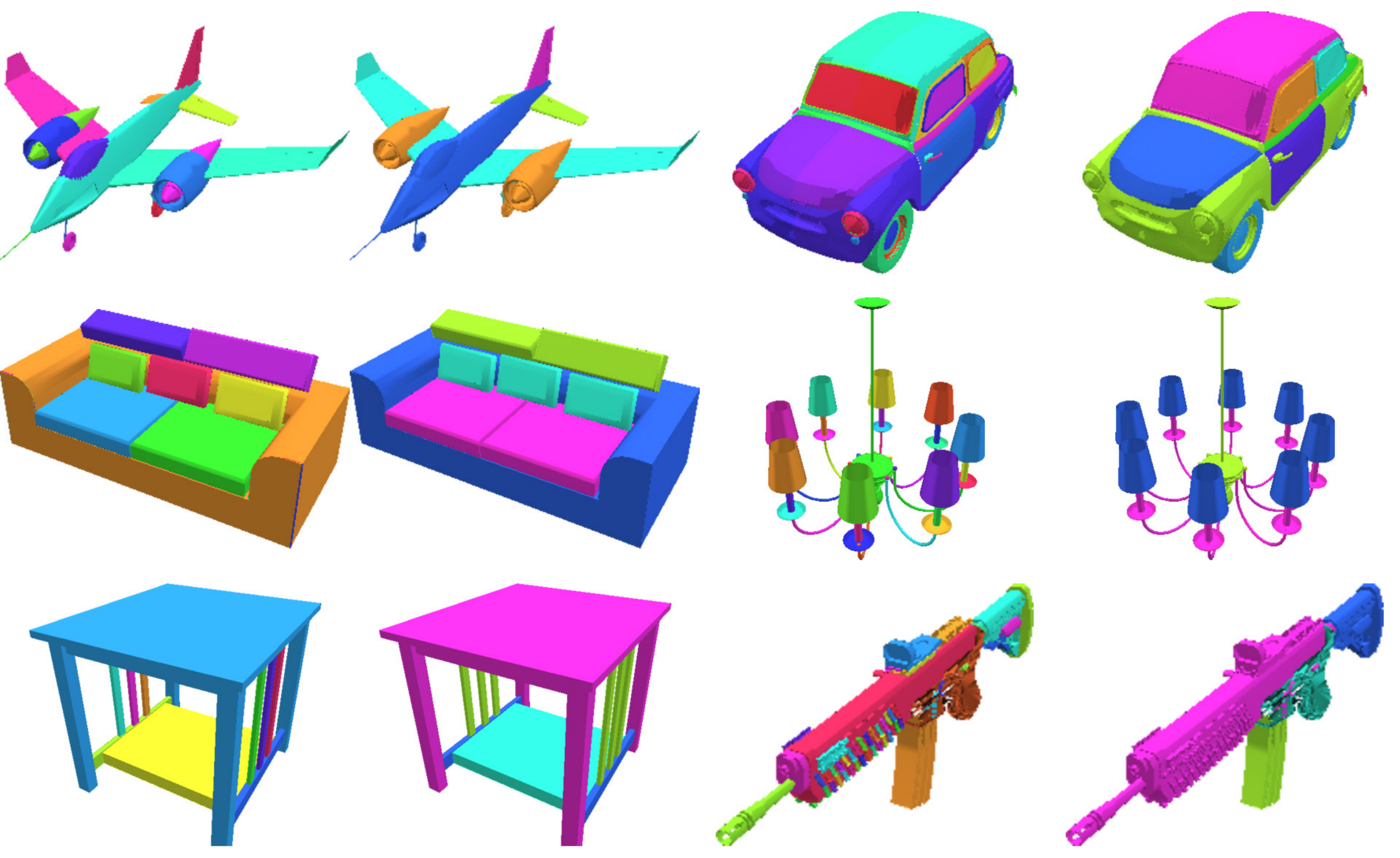}
\caption{\rev{This figure illustrates our pre-processing step, where the image on the left depicts input connected components, and image on the right are the nodes are the final components (after small, overlapping, and repetitive elements are grouped together).}}
\label{fig:component_examples}
\end{figure}

Given a database of shapes the goal of this step is to decompose them into components and construct contact graphs over the components. \rev{We can partition this graph in various ways to create training pairs of a partial assembly (connected subgraph) and its complements (nodes adjacent to the subgraph).}
This step has loose requirements, since subsequent steps do not require these components to be consistent or labeled. That said, it is desirable for these components to have non-negligible size so that adding them makes a visible difference to the assembly, and have their boundaries roughly align with geometric features to avoid visual artifacts in stitching the parts together. Larger components also aid in learning a more meaningful and discriminative embedding space. Thus, we start with an over-segmentation where each component has reasonable boundaries and then iteratively merge small components. 
While we could use an automatic segmentation algorithm such as randomized cuts~\cite{Golovinskiy:2008} to produce the initial components, we found ShapeNet models are already represented by scene graphs where leaf geometry nodes provide reasonable components with minimal post-processing.  

\rev{We first construct an initial contact graph by creating an edge between any two components such that the minimum distance between them is less than $\tau_\text{proximity}=0.05$ of their radius. We then choose a set of nodes to merge into a single component based on three criteria: size, amount of overlap, and similarity. Specifically, any component with PCA-aligned bounding box diagonal below $\tau_\text{size}=0.2$ of mesh radius is merged to its largest neighbor. Also, overlapping components with \emph{directional Hausdorff} distance below $\tau_\text{Hausdorff tol}=0.05$ (in either direction) are merged into the same group. Finally, identical components that share the same geometry in the scene graph or with identical top/front/side grayscale renderings are treated as a single component. The last merge favors placing all symmetric parts at once, which we found to be more time effective from the user perspective (e.g., think of placing every slat separately to form a back). The output of these merges is a new contact graph and we synthesize training pairs by partitioning this graph in different ways (\Sec{joint_training}). We only use graphs that have at most $N_\text{max CC} = 8$ components during training. 
We demonstrate the effect of these pre-processing steps in Figure~\ref{fig:component_examples} and statistics over our training data in Figure~\ref{fig:data_stats}.}



For our retrieval and placement networks we represent our components with $N_\text{points}=1000$ randomly sampled points re-centered at the origin.

\begin{figure}[t]
\centering
\includegraphics[width=0.5\textwidth]{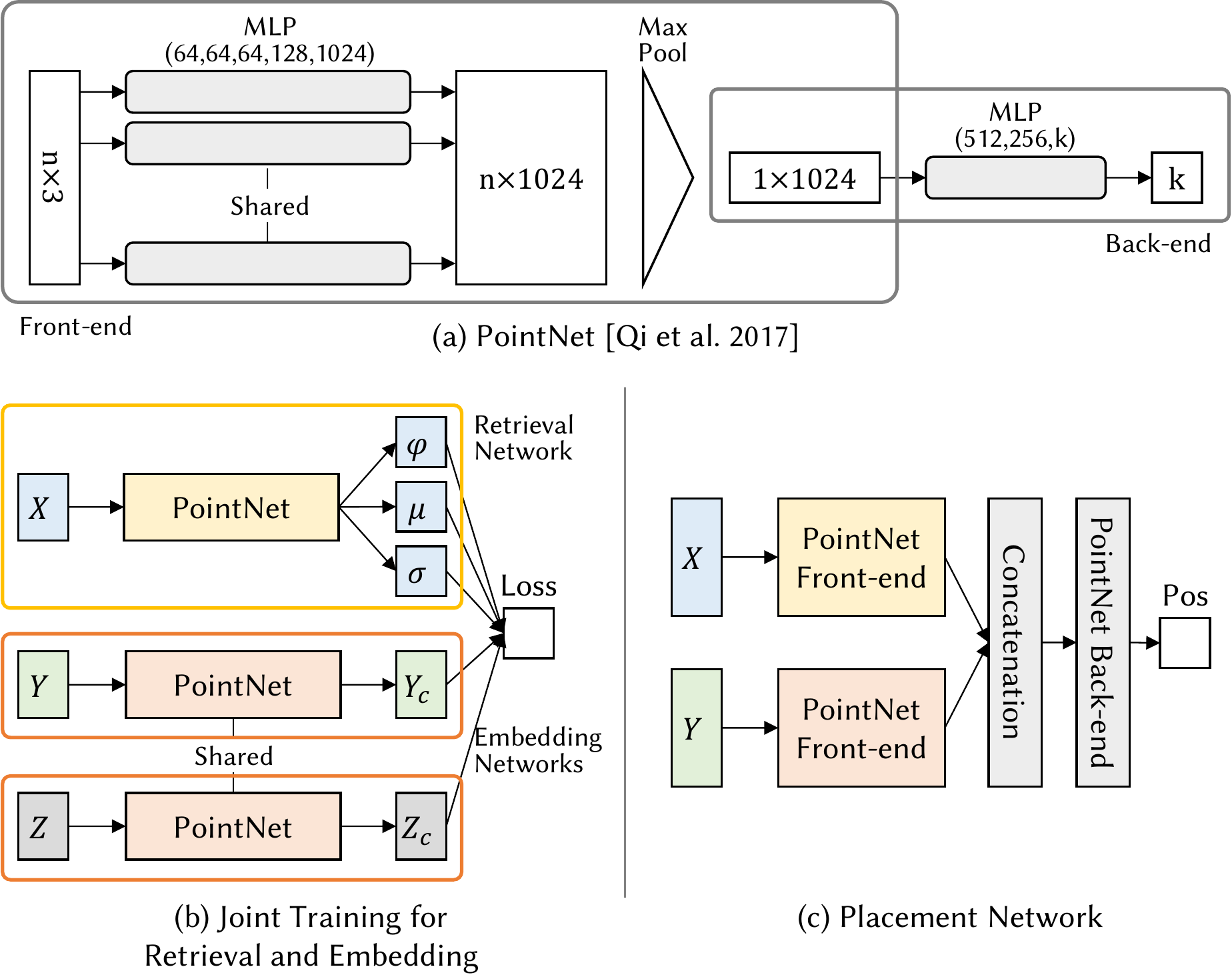}
\caption{\rev{Neural networks architectures. (a) PointNet \cite{Qi:2017} (provided for completeness). The numbers in MLP (Multi-Layer Perception) are layer sizes. In our use, we omit spatial transformer networks and assume that orientations are pre-aligned (\Sec{placement_network}). (b) the joint training framework of both retrieval and embedding networks. (c) the placement network.}}
\label{fig:network}
\end{figure}

\section{Method}
\label{sec:method}

The input to our method is a partial shape and the output are several component proposals selected from the database that can be added next. 
We design several neural networks to facilitate the proposal: a retrieval network $g$, an embedding network $f$, and a placement network $h$.
\rev{While our networks can be re-targeted to deal with any 3D shape representation such as voxel grids or multi-view projections we chose to represent all input shapes with point clouds~\cite{Qi:2017} which are versatile representations that can be used on a wide range of geometries.}
%

To index parts we build an embedding space for all components, where interchangeable and stylistically compatible components are embedded nearby. We represent this space with a neural network $f$ that takes part geometry and maps it to low-dimensional vector. The retrieval network and embedding network are tightly coupled. The retrieval network $g$ takes geometry of a partial query as an input, and outputs a probability distribution over the low-dimensional embedding learned by $f$ (see \Fig{overview}). 

A good embedding needs to provide a space that is easy to represent with the output of the retrieval network. Thus, we jointly train both networks with triplets: a partial shape, one of its complements, and a non-complementing part. We then separately train a placement network $h$ that takes geometry of the query shape and a retrieved complement and outputs placement coordinates for the component.

\subsection{Retrieval Network}
\label{sec:retrieval_network}
Given a partially assembled shape our goal is to retrieve a set of complementary parts. 

Our input partial assembly $X$ is represented as a point cloud of $N_\text{points}$ points. Note that any partial shape $X$ can have several complementary parts, thus instead of predicting a unique coordinate, $Y_c$, we predict a conditional distribution over the embedded space, $P(Y_c | X)$. We model $P(Y_c|X)$ as a mixture of Gaussians, defined on some $D-$dimensional embedding space, i.e., $Y_c \in \mathbb{R}^D$ (where $D=50$ in all experiments). 

We predict the distribution by mixture density network (MDN) \cite{Bishop:1994}, which essentially predicts the parameters of the Gaussian mixture. 
For the mixture of Gaussians, we use $N_\text{GM}=N_\text{max CC}$ modes in our model, set to maximal number of connected components, and represent each $k^\text{th}$ Gaussian with a weight $\phi_k \in \mathbb{R}$, a mean $\mu_k\in \mathbb{R}^D$, and a standard deviation $\sigma_k \in \mathbb{R}^D$. 

To take unordered points as input we use the PointNet network~\cite{Qi:2017} as the backbone structure, which leverages symmetric (order-independent) functions to map points to categories \rev{(see \Fig{network}a)}. To predict probability distribution over the embedded space we replace classification output layers with parameters of Gaussian mixture model: $g(X) = \{\phi_k(X), \mu_k(X), \sigma_k(X)\}_{k=1..N_\text{GM}}$, where $g$ is a PointNet architecture. Each of weights, means, and standard deviations are mapped from the feature of the input with a single fully connected layer for each with different activations: softmax for weights to make sum one, exponential for variances to constrain them to be positive, and linear for means.

It is worth mentioning that, the modeling of a conditional distribution over neural network output is an active research field recently, and our choice of MDN as the tool is mainly due to its significantly better performance to capture multiple modes. In principle, recent techniques such as conditional GAN \cite{Mirza:2014} or conditional VAE \cite{Kingma:2014,Sohn:2015} can also be used here; however, it is well-known that these approaches are still unapt to capture multiple modes well, suffering from a phenomenon known as mode collapse.

\begin{figure}[t]
\centering
\includegraphics[width=0.8\linewidth]{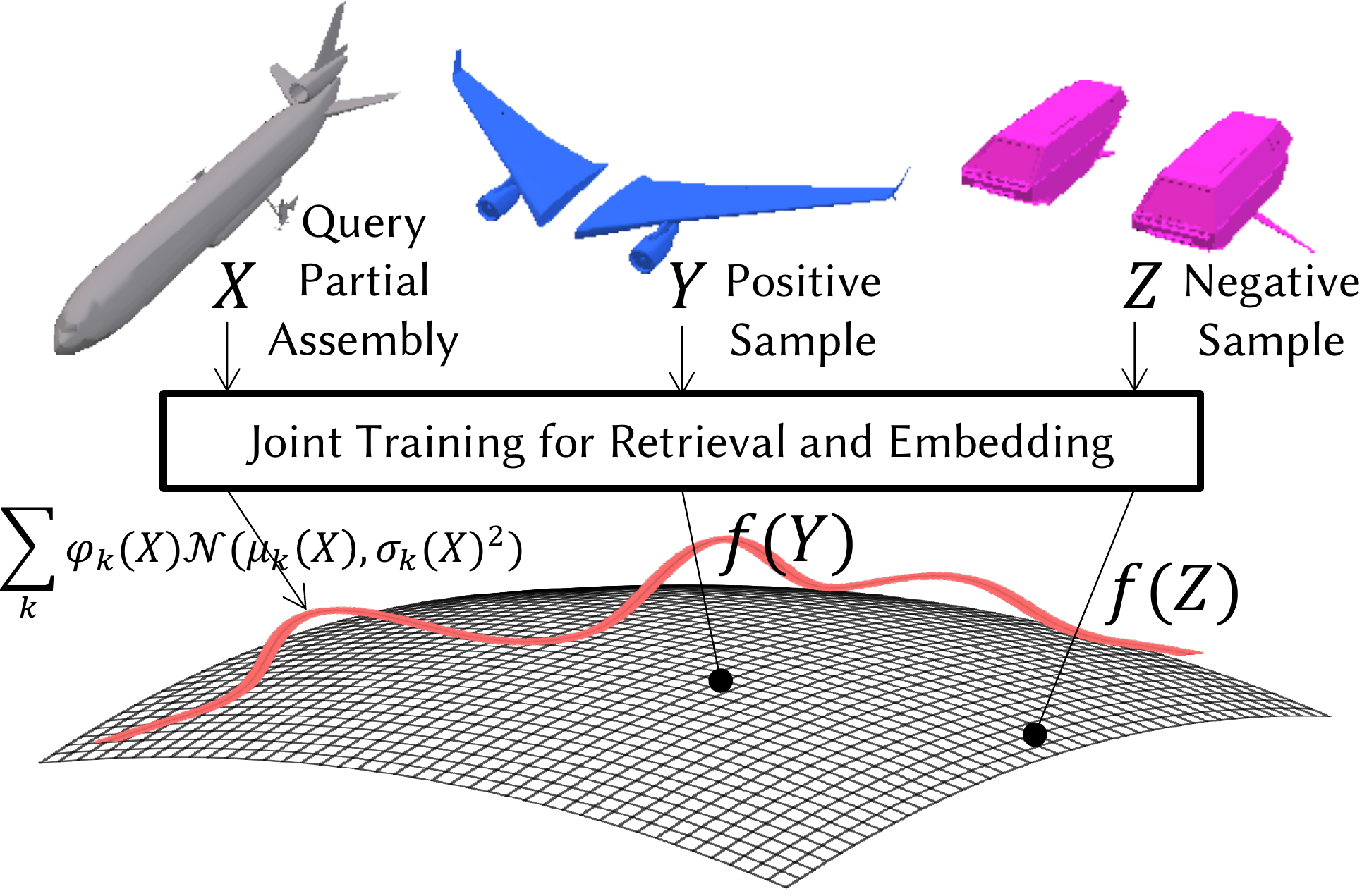}
\caption{A triplet of the query partial assembly $X$, a positive sample $Y$, and a negative sample $Z$ becomes an instance of the training data. Function $f(\cdot)$ maps $Y$ and $Z$ to single points on the embedding space, and function $g(\cdot)$ generates Gaussian mixture distribution on the space from $X$. See \Sec{joint_training} for details.}
\label{fig:training}
\end{figure}

\subsection{Embedding Network}
\label{sec:embedding_network}
The next step is to design the embedding network $f$ that takes a shape and maps it to the $D-$dimensional embedding space. 
Since the retrieval network works by predicting a coordinate in the space and selecting
candidates by proximity search, nearby components have to be interchangeable
when they are added to some partial object assemblies.
A naive approach would be to use some fixed embedding (e.g., PCA) based on any shape features (e.g., deep learned classification features~\cite{Su:2015}). The disadvantage of this approach is that embedding is created independently from the prediction network $g$, so we cannot expect complementary parts to be captured well with the Gaussian Mixture model. 
%
%
Thus, we propose to learn the embedding space jointly with the network $g$. To do this we use a PointNet architecture to represent function $f(Y)$ that maps the point cloud of a component $Y$ to its embedding coordinates $Y_c$.  Learning the embedding function $f$ enables us to create an embedding space that tightly clusters candidate complements that share stylistic and functional attributes. 

\subsection{Joint Training for Retrieval and Embedding}
We now describe how to jointly train the retrieval and embedding networks using our pre-processed dataset.
\label{sec:joint_training}

{\noindent \bfseries  Loss function.}
Our loss is a triplet contrastive loss. Given some positive example of a partial assembly $X$ and its complementing part $Y$, we need to define an appropriate loss function to be able to learn optimal parameters for networks $f$ and $g$. We define it as a negative log likelihood that $Y$ is sampled from the probability distribution predicted by $g(X)$, $P(Y | X)$:
\begin{align}
E(X, Y) &= -\log \sum_{k=1}^{N_\text{GM}} \phi_k(X) \mathcal{N}(f(Y) \,|\, \mu_k(X), \sigma_k(X)^2 ).
\label{eq:loss_joint}
\end{align}
\rev{See Appendix for an expanded form.} Directly optimizing for parameters of $g$ and $f$ with respect to Equation~\ref{eq:loss_joint}, however, would collapse the embedding space to a single point, i.e., the optimal value is attained when $f$ contracts to a single point~\cite{Hadsell:2006}. 
Thus, we introduce a negative example, component $Z$ that does not complement $X$, to avoid the contraction of $f$.
We now use the triplet $(X,Y,Z)$ to define a contrastive loss~\cite{Chechik:2010}:
\begin{align}
E(X, Y, Z) &= \max \{m + E(X, Y) - E(X, Z), 0 \},
\label{eq:loss_constrasive}
\end{align}
where $m=10$ is a constant margin set for all experiments.

\Fig{network}b shows the final version of the network for the component embedding with the contrastive loss. The subnetworks processing $X$, $Y$, and $Z$ have the same PointNet structure, but only the subnetworks of $Y$ and $Z$ share parameters. 

\vspace{0.1cm}
{\noindent \bfseries  Training.}
To generate the training triplet $(X, Y, Z)$ we use the components in the pre-processed contact graphs described in Section~\ref{sec:data_preprocessing}. We first pick a random shape. Suppose its contact graph has $n$ nodes, we then pick a random value $r \in [1,n]$ and create a random subgraph with $r$ nodes. To do that we pick a random node and iteratively add a random unvisited adjacent node until we create a connected subgraph graph of size $r$. We sample $N_\text{points}$ on the included components to obtain $X$ (note that these points are defined in global coordinate system of the object). We then pick a random unvisited component that is adjacent to the selected subgraph $X$ to define $Y$, and a random non-adjacent component (including components from other shapes) to define $Z$ (note that $Y, Z$ are represented by $N_\text{points}$ centered at the origin). 

We train the retrieval network for $2000$ epochs with batch size $32$.  We use ADAM optimizer \cite{Kingma:2014b} with $1.0E-3$ initial learning rate and 0.8 decay rate for 50k decay steps. Each epoch iterates all 3D models in the training set while randomly sampling the query subgraph $X$, and positive/negative components $Y$ and $Z$. In MDN, the standard deviations ${\sigma_k(X)}$ easily diverges to the $\inf$ since this leads to $-\inf$ loss. Hence, we set the upper limit of  ${\sigma_k(X)}$ to 0.05.

\subsection{Placement Network}
\label{sec:placement_network}
The retrieval network predicts a probability on the embedding space. We can accordingly propose new components to be selected for interactive or fully-automatic model design. Suppose that $Y$ is the selected new component given a partial object $X$. The placement network $h$ predicts 3D coordinates for the component $Y_p=h(X,Y)$. We assume that only translation needs to be predicted and orient $Y$ the same way as it was oriented in the source shape. We use two independent PointNet networks to analyze point clouds $X$ and $Y$, concatenate the features from these two networks, and add multilayer perceptron layers to obtain 3D coordinates $Y_p$ (\Fig{network}c).  We use the same training data samples $(X,Y)$ as in training the retrieval network.


\section{Results}
\label{sec:experiments}

We demonstrate interactive and automatic modeling tools that can leverage our method. We also quantitatively evaluate our method and compare to the state-of-the-art alternatives. 

\begin{figure}
\centering
\includegraphics[width=\linewidth]{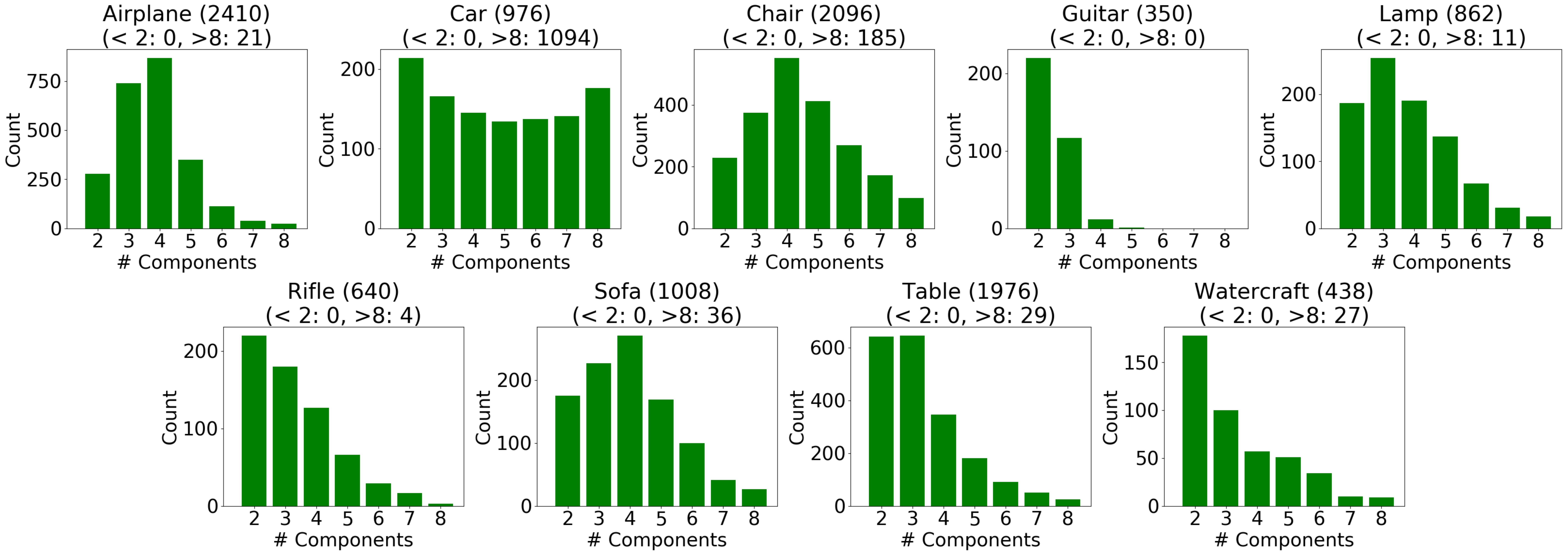}
\caption{Histograms of numbers of components. \rev{The number next to the category name is the total number of models we used in experiments (including train/test), and the next line is the numbers of discarded models which have less than two or greater than eight components.}}
\label{fig:data_stats}
\end{figure}

\vspace{0.1cm}
{\noindent \bfseries Dataset.}
We test our method with 9 categories from ShapeNet repository~\cite{Chang:2015}: Airplane, Car, Chair, Guitar, Lamp, Rifle, Sofa, Table, and Watercraft. We picked diverse categories with interesting part structures and enough instances to provide useful training data. Our pre-processing produces a few components per shape and we disregard shapes that have \rev{only 1 or more than 8 components} (see \Fig{data_stats} for details).
An interactive modeling or a shape synthesis tool performs the best if they leverage the entire dataset, so the qualitative results provided in Section~\ref{sec:geometricmodel} are trained on the entire dataset. For quantitative evaluations and comparisons for various retrieval and placement algorithms we randomly split every category into 80\% for training and 20\% for test sets and report quantitative results and qualitative comparisons on test sets only (Section~\ref{sec:evaluation}). 

\vspace{0.5cm}

\subsection{Assembly-Based Geometric Modeling}
\label{sec:geometricmodel}
We first evaluate our method qualitatively for interactive and automatic shape modeling. 

\vspace{0.1cm}
{\noindent \bfseries Interactive Modeling.}
We use our retrieval and placement networks in an interactive modeling interface.
Given a partial assembly, our algorithm first proposes a set of possible components by sampling from the conditional probability distribution predicted by the retrieval network and shows the candidates in our UI.
\setlength{\columnsep}{10pt}
\begin{wrapfigure}{r}{0.20\textwidth}
  \begin{center}
    \includegraphics[width=0.20\textwidth]{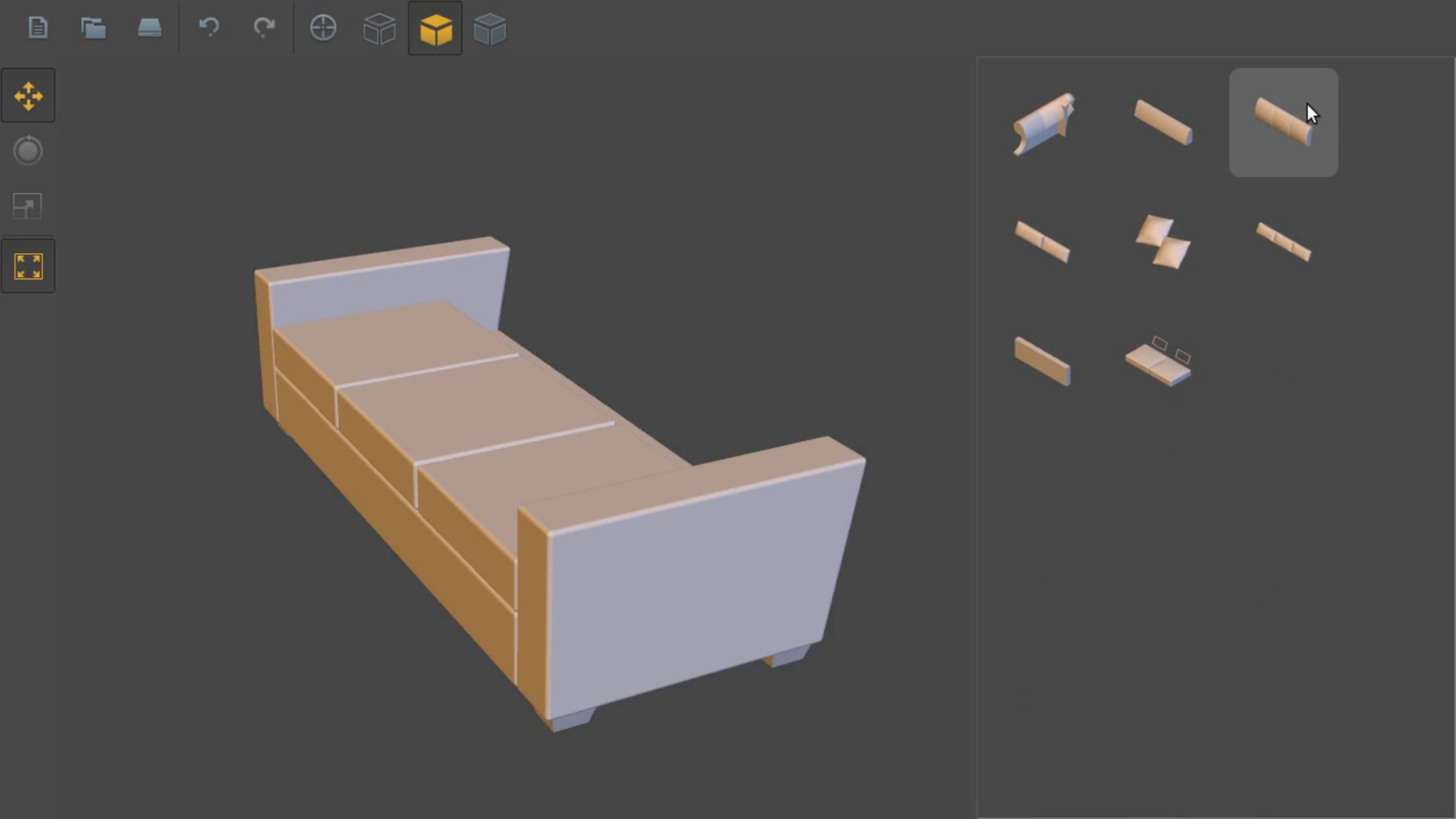}
  \end{center}
\end{wrapfigure}
Then, the user selects a desired complementary component, and the algorithm predicts the location for it via the placement network. The new shape is synthesized for the user, and the next component is proposed.  Refer to the supplemental video for several interactive sessions.

\begin{figure*}[t]
\centering
\includegraphics[width=\linewidth]{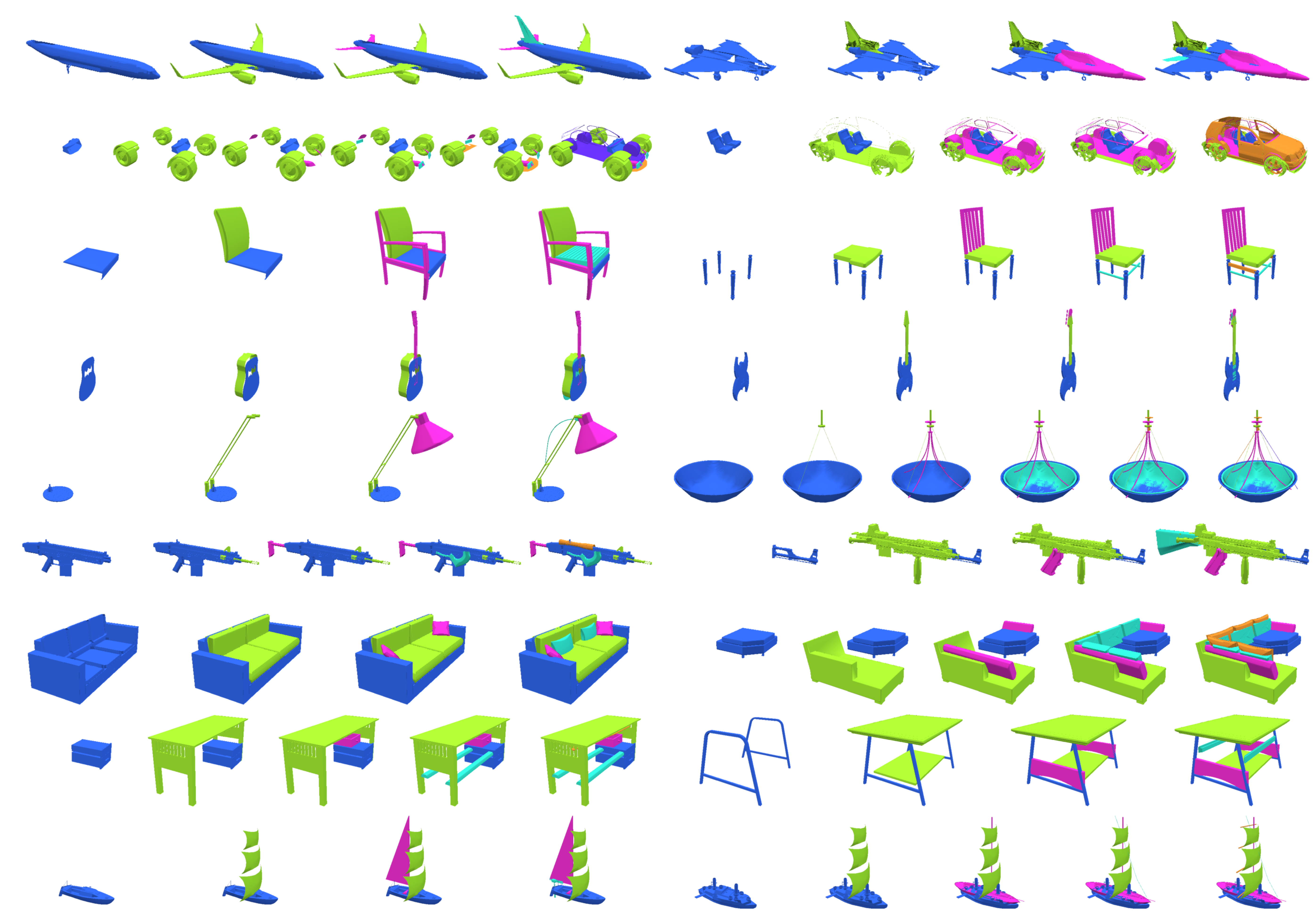}
\caption{Automatic iterative assembly results from a single component. Small component not typically labeled with semantics in the shape database (e.g. as slats between chair/table legs, pillows on sofas, cords in lamps/watercrafts) are appropriately retrieved and placed.}
\label{fig:automatic_assembly}
\end{figure*}

\begin{figure*}[t]
\centering
\includegraphics[width=\linewidth]{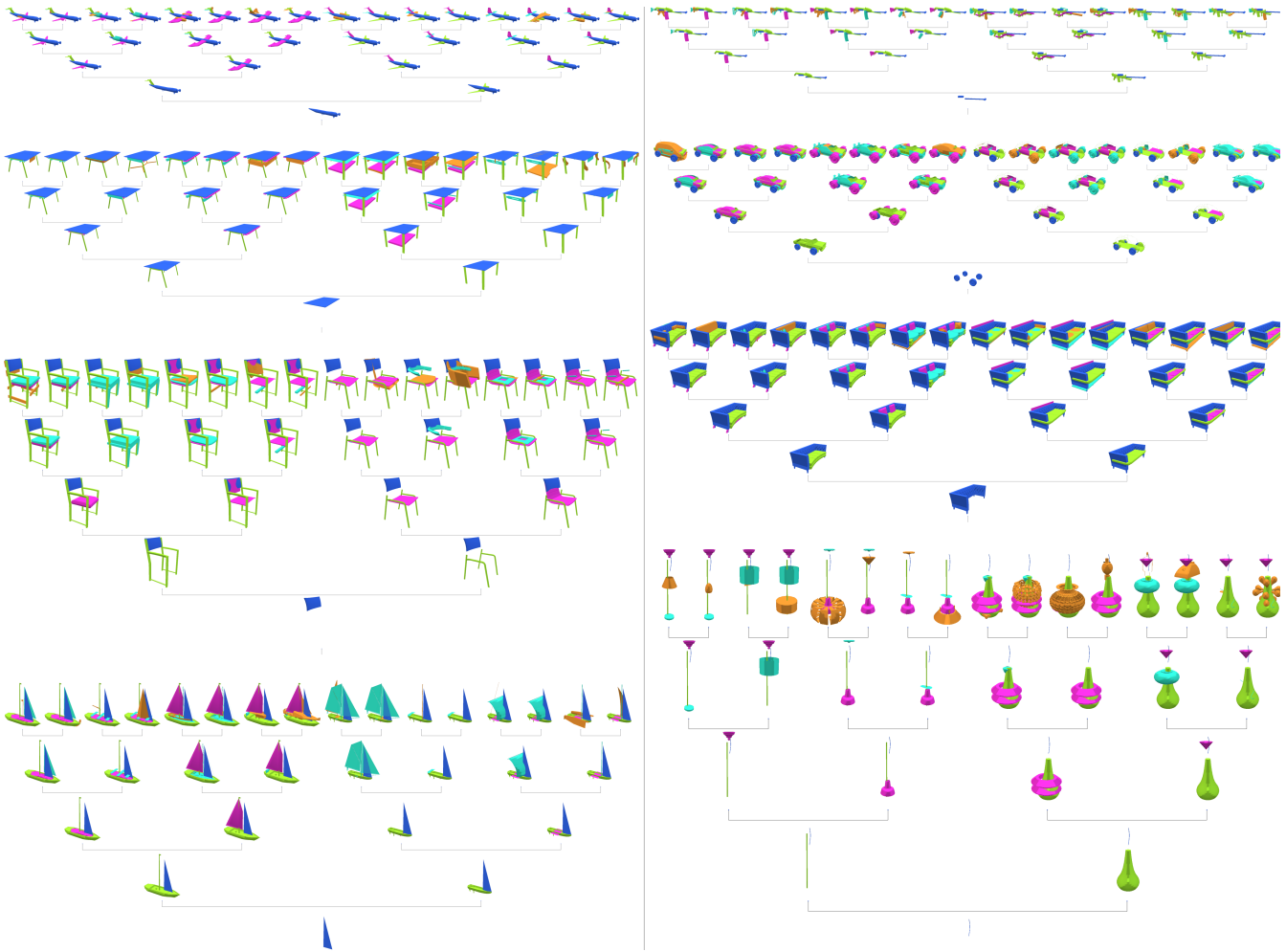}
\caption{Automatic iterative assembly with two different random choices at every time. From the initial component at the bottom, various objects are synthesized by assembling different components.}
\label{fig:assembly_trees}
\end{figure*}

\vspace{0.1cm}
{\noindent \bfseries Automatic Shape Synthesis.}
Our method can also be used to facilitate a fully automatic shape synthesis that is able to generate diverse designs. We simply start with a random component, and iteratively add a component by sampling from the predicted distribution. Figure~\ref{fig:automatic_assembly} shows the evolution of the model when the component with maximal probability is added at every iteration. The retrieval network successfully finds new components that are missing in the query and can be connected to the given partial assembly. At each step, one can also make various decisions by taking different components from the sampling, so in Figure~\ref{fig:assembly_trees} we show a binary tree of possibilities. Note that in various content creation scenarios one can use this to control the complexity of resulting models (based on the depth of the tree) and diversity of the resulting models (based on the breadth of the tree).

\subsection{Quantitative Evaluations and Comparisons}
\label{sec:evaluation}
Evaluating an assembly-based geometric modeling tool is a challenging problem since there is no well-established protocol. In particular, evaluating the whole end-to-end object design process relies on subjective user evaluations which is prone to bias (e.g., the modeling task can be geared to favor a particular method). We thus propose a benchmark that evaluates various aspects of our core contributions: complement retrieval, part embedding, and part placement. 

Evaluating whether retrieved parts are compatible with the query partial shape is not a trivial task, since compatibility is a subjective criterion. We propose to evaluate functional, geometric, and stylistic compatibilities with separate metrics outlined in the following paragraphs. For each criterion, we compare our result to \rev{a random suggestion baseline} and two state-of-the-art alternatives. First is the method of Chaudhuri and Koltun~\shortcite{Chaudhuri:2010} (CK10) that also does not require a database of labeled parts and thus is directly comparable with our input. Their method operates in two steps: they find shapes that are similar to the query using global shape descriptors, and then pick components in the retrieved shapes that are dissimilar from the components in the query. Second, to test the value of our embedding, we replace  our joint training of embedding network $f$ and retrieval networks $g$ with a fixed embedding space from MVCNN~\cite{Su:2015} (i.e., only $g$ is trained in this case). More specifically, we extract the last layer of MVCNN and use PCA to project it to a 50-dimensional space.

We also evaluate the part placement network and present quantitative results to enable future comparisons. 

\vspace{0.1cm}
{\noindent \bfseries Functional compatibility.}
To answer the question whether a retrieved part is functionally compatible with the partial query, we rely on existing segmentation benchmark with part labels that refer to their functionality (e.g., an airplane can include four functional parts: a body, an engine, a tail, and wings).  We then remove a single part from the query shape and evaluate how many of the retrieved components have correct labels. 

In this experiment we use part labels in the ShapeNet dataset~\cite{yi:2016}. Since this dataset provides per-point labels rather than isolated components we first label connected components and group them into bigger parts. In particular, we use majority voting to label each connected component, and then group all components with the same label into a single part. We disregard shapes if they have the final labeled part covering less than 80\% of the labeled points in the dataset (we use 3396 out of 8670 models in this evaluation). 
This provides us with a database of shapes that are decomposed into consistent semantic parts. Note that this is very different from our training components obtained after database preprocessing in Section~\ref{sec:data_preprocessing} which are inconsistent and unlabeled. We do not use these labeled components for training, but only use them to create the query shapes and component database in this experiment. 

In this experiment, we report numbers on 6 categories for which ShapeNet has part annotations out of the 9 categories we tested. We generate 100 queries for each category. For each query, we exclude a randomly chosen part for each shape and measure the mean average precision scores (mAPs) of the top 5 retrieval results (where a result is considered to be correct if the part label matches the label of excluded part).  We present quantitative results in Table~\ref{tbl:part_label_map} demonstrating that our method outperforms CK10 on all categories, except guitars, where both methods perform well due to very regular structure of the shape. 
%

\begin{table}[ht]
{\small
\setlength\tabcolsep{4pt}
\begin{tabular}{c|*{6}c|c}
\toprule
 Category  &  Plane  &  Car  &  Chair  &  Guitar  &  Lamp  &  Table  &  Mean  \\
\midrule
  Random & 0.41 & 0.40 & 0.42 & 0.51 & 0.49 & 0.63 & 0.48 \\
  CK10 & 0.37 & 0.39 & 0.37 & \textbf{1.00} & 0.52 & 0.64 & 0.55 \\
  Ours (MVCNN)  & 0.70 & \textbf{0.71} & 0.80 & 0.93 & 0.74 & 0.88 & 0.79 \\
  Ours (Joint) & \textbf{0.89} & 0.55 & \textbf{0.86} & 0.95 & \textbf{0.73} & \textbf{0.91} & \textbf{0.81} \\
\bottomrule
\end{tabular}
}
\caption{Evaluating functional labels of retrieved parts, this table reports mean average precision for top 5 retrievals across different methods and categories.  }
\label{tbl:part_label_map}
\vspace{-0.5cm}
\end{table}

\vspace{0.1cm}
{\noindent \bfseries Geometric compatibility.}
Even if a retrieved part has correct label, it might not fit well with the query shape. While it is hard to evaluate geometric compatibility, we resort to comparing geometry of the retrieved part and the original part that was excluded from the assembly. We use the same experimental setup as in evaluating functional compatibility, but measure the average Hausdorff distance between the original and top 5 retrieved parts (\Tbl{hausdorff_distances}). The distances are relative to the shape radius which is scaled to 1. Note that our method returns parts that are more similar to the original complement than parts returned by CK10. Similar to functional compatibility metric, we found that our method performed slightly worse on guitars, where global shape descriptors might be mostly appropriate to capture guitar shapes. 
%

\begin{table}[ht]
{\small
\setlength\tabcolsep{4pt}
\begin{tabular}{c|*{6}c|c}
\toprule
 Category  &  Plane  &  Car  &  Chair  &  Guitar  &  Lamp  &  Table  &  Mean  \\
\midrule
  Random & 0.27 & 0.24 & 0.27 & 0.14 &0.30 & 0.36 & 0.26 \\
  CK10  & 0.23 & 0.21 & 0.27 & \textbf{0.04} & 0.25 & 0.34 & 0.23 \\
  Ours (MVCNN) & 0.15 & \textbf{0.15} & 0.19 & 0.05 & 0.22 & 0.27 & 0.17 \\
  Ours (Joint) & \textbf{0.11} & 0.21 & \textbf{0.16} & 0.05 & \textbf{0.21} & \textbf{0.24} & \textbf{0.16} \\
\bottomrule
\end{tabular}
}
\caption{Evaluating geometric compatibility, this table report average Hausdorff distances of top-5 retrievals with the ground truth missing parts. Note that all models are normalized to have unit radius.}
\label{tbl:hausdorff_distances}
\vspace{-0.5cm}
\end{table}

\vspace{0.1cm}
{\noindent \bfseries Style compatibility.}
Our next goal is to evaluate whether the retrieved component is compatible to the query in style. While this is not a well-formulated problem, for the purpose of evaluation we use ShapeNet \emph{taxonomy} to reason about finer-grained classes, since such fine-grained classes are often defined by style (e.g., chair's find-grained classes are club chair, straight chair, lounge chair, recliner, etc). In particular, we consider a retrieved part to be accurate if it is from a shape in the same fine-grained class. While this is generally an imperfect measure (e.g., car wheels are interchangeable between convertibles and jeeps and table legs are interchangeable between rectangular tables and round tables, even though global fine-grained labels are different), we found that this measure does correlate with style compatibility of query and target shape, and nicely complements other measures. 
We evaluate style compatibility on 6 categories that have various fine-grained classes.

We test two extremes: only one part is missing from the query, and only one part is present in the query.
One issue is that the fine-grained categories of ShapeNet models do have overlaps, i.e., one model may have multiple fine-grained class labels (e.g. club chair and armchair), and thus the subclasses are considered to be matched if there is any overlap between the sets of subclasses of the query and the retrieval results. 

\Tbl{texonomy_map} shows mAPs of top 5 retrieval results. Our method performs better when only one part is given, suggesting that it can capture stylistic compatibility with very little information. However, we found that global shape descriptors used by CK10 perform better at retrieving stylistically similar shape when the query has almost the complete shape. It is worth noting that our method has more disadvantage under this metric. We introduce an embedding space and in it components from different fine-grained categories may be near each other (the car wheel example before). However, the CK10 approach tends to find a shape very similar to the query in the all-expect-one setting, with a high chance to be from the same fine-grained category. 
Note that this metric only reasons about style of the global shape, even if individual retrieved parts look identical.
%

\begin{table}
{\small
\setlength\tabcolsep{4pt}
\begin{tabular}{c|*{6}c|c}
\toprule
 Category & Plane & Car & Chair & Sofa & Table & Ship & Mean \\
\midrule
   Random & 0.65 & 0.28 & 0.50 & 0.38 & 0.37 & 0.50 & 0.50 \\
\midrule
   \multicolumn{8}{c}{All except one } \\
\midrule
   CK10 & \textbf{0.79} & \textbf{0.80} & \textbf{0.83} & \textbf{0.84} & \textbf{0.71} & \textbf{0.75} & \textbf{0.79} \\
   Ours (MVCNN) & 0.76 & 0.32 & 0.67 & 0.52 & 0.47 & 0.56 & 0.60 \\
   Ours (Joint) & 0.78 & 0.34 & 0.65 & 0.67 & 0.49 & 0.59 & 0.62 \\
\midrule
   \multicolumn{8}{c}{Single } \\
\midrule
   CK10 & 0.51 & \textbf{0.44} & 0.52 & 0.56 & 0.44 & 0.54 & 0.49 \\
   Ours (MVCNN) & 0.71 & 0.29 & 0.64 & 0.54 & 0.46 & 0.53 & 0.58 \\
   Ours (Joint) & \textbf{0.72} & 0.26 & \textbf{0.64} & \textbf{0.59} & \textbf{0.52} & \textbf{0.59} & \textbf{0.59} \\
\bottomrule
\end{tabular}
}
\caption{Evaluating style compatibility, this table reports mAP for fine-grained style categories of retrieved components.}
\label{tbl:texonomy_map}
\vspace{-0.5cm}
\end{table}

\begin{figure*}
\centering
\includegraphics[width=\textwidth]{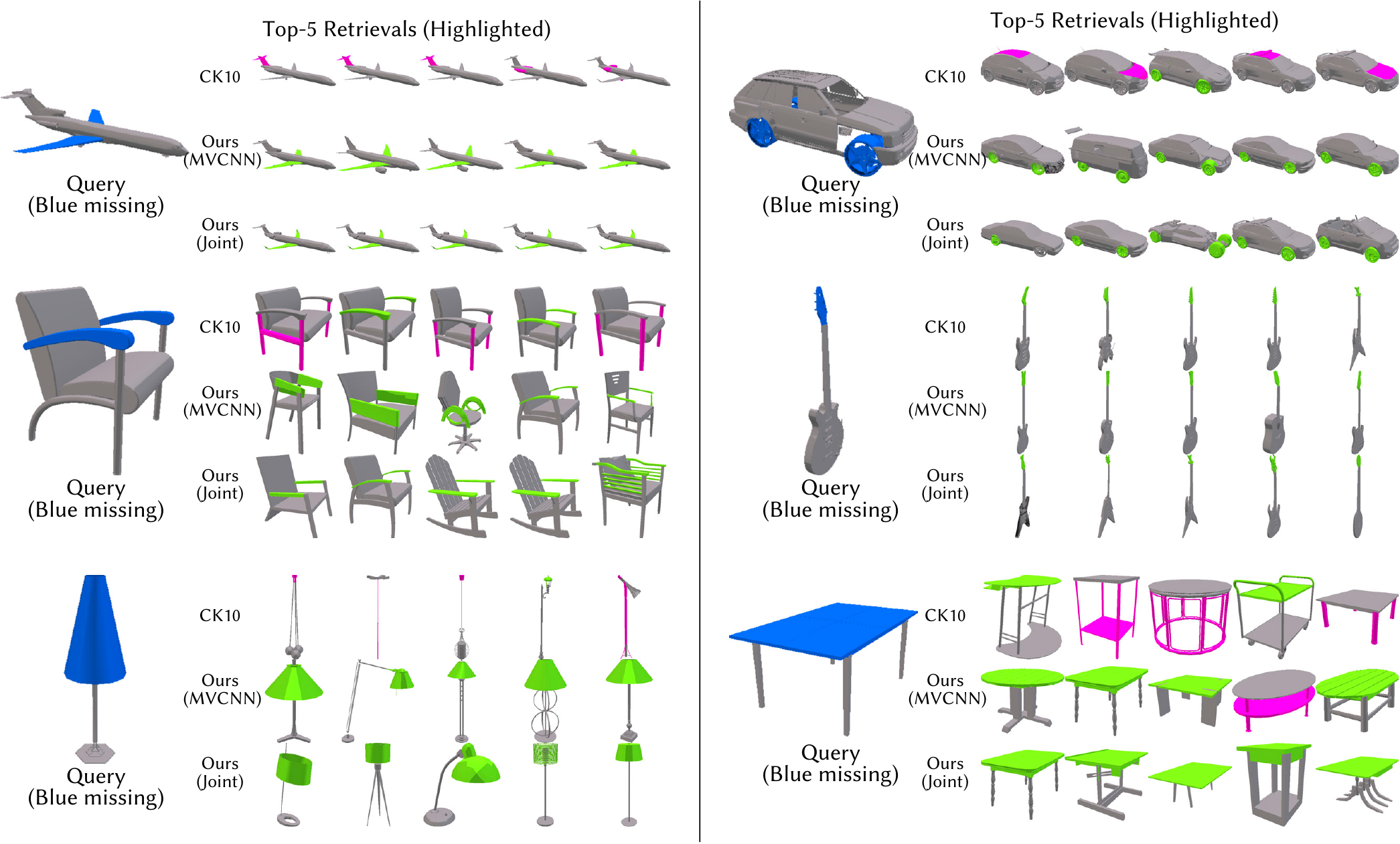}
\caption{Part retrieval comparisons. The left is query shape with a missing component (blue is missing). The three rows on the right are top-5 retrieval results of CK10, ours with MVCNN embedding, and ours with joint embedding, respectively. The retrieved components are highlighted with light green and pink; light greens are correct, and pinks are wrong parts.}
\label{fig:comparisons}
\end{figure*}

\textbf{Comparison to Chaudhuri and Koltun~\shortcite{Chaudhuri:2010} (CK10).}
Previously we mentioned that CK10 relies on hand-crafted global shape descriptors to retrieve the most similar shape, and then propose parts that are dissimilar to the parts in the query. In contrast, our method learns to predict the descriptor of a complement from the query, which is a more direct method. We also use neural networks for this task, which enable our approach to leverage large datasets as they become available. We demonstrate some qualitative results in \Fig{comparisons}.

\vspace{0.1cm}
{\noindent \bfseries Effect of learning the embedding.}
We also evaluate the influence of using fixed embedding space instead of learning the embedding. 
In particular, we pick MVCNN descriptor as one of the state-of-the-art deep learned shape descriptors and train only the retrieval network $g$, while $f$ is prescribed by the PCA transformation of the MVCNN descriptor in the training data.

Evaluated by part label and style prediction metrics, \Tbl{part_label_map} and \Tbl{texonomy_map} show that our method based on the learned embedding works on par or slightly better than the feature space from MVCNN.  
 
\begin{figure*}
\centering
\includegraphics[width=\textwidth]{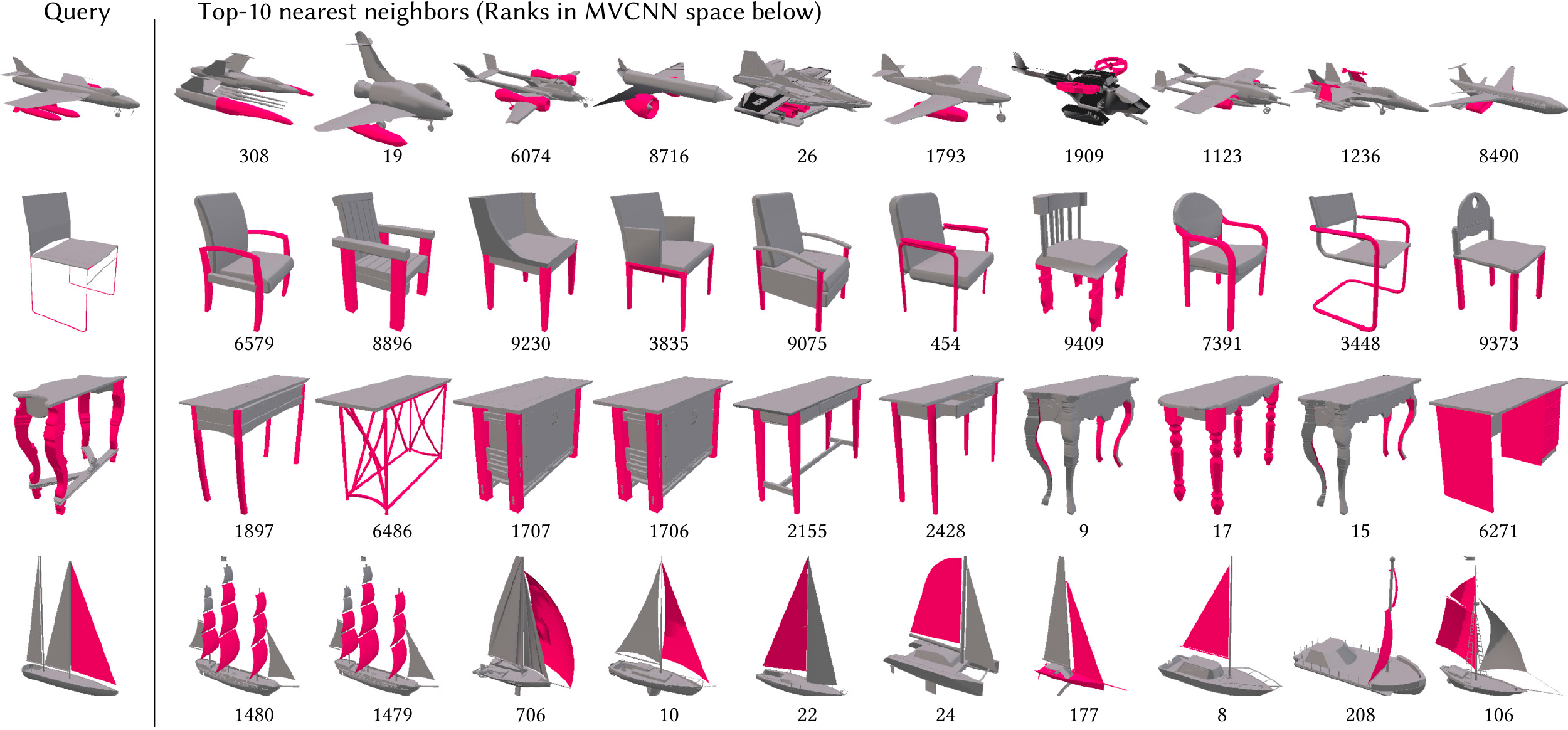}
\caption{Neighborhood in the embedding space learned by our embedding network. Top-10 nearest neighbors from the query with the nearest neighbor ranks in the MVCNN embedding space below.}
\label{fig:neighbors}
\end{figure*}

Qualitatively, we find that the learned embedding often exhibits larger diversity orthogonal to the appearance similarity captured by MVCNN. In \Fig{neighbors}, we visualize our learned embedding space and observe such diversity by our learned embedding. For example, the returned table legs in the third row differ greatly in shape but are all reasonable components to be added to the partial assembly.   

\vspace{0.1cm}
{\noindent \bfseries Part placement evaluation.}
A unique advantage of our method over CK10 and existing approaches is its ability to predict part placement purely from the geometry of the new component. 
\Tbl{placement_error} shows the placement error on both training and testing data, where the error is measured relative to shapes with the unit radius. Note that all models in the database are normalized to have unit radius from the bounding box center. In all categories, our placement network predicts positions with reasonably small errors. 
%

\begin{table}
{\small
\setlength\tabcolsep{2pt}
\begin{tabular}{c|*{9}c|c}
\toprule
 Category & Plane & Car & Chair & Guitar & Lamp & Rifle & Sofa & Table & Ship & Mean \\
\midrule
  Train Err & 0.02 & 0.03 & 0.06 & 0.02 & 0.04 & 0.02 & 0.04 & 0.06 & 0.03 & 0.04 \\
  Test Err & 0.06 & 0.13 & 0.13 & 0.03 & 0.20 & 0.14 & 0.12 & 0.15 & 0.19 & 0.12 \\
\bottomrule
\end{tabular}
}
\caption{Placement Network Error. Note that all models are normalized to have unit radius.}
\label{tbl:placement_error}
\vspace{-0.5cm}
\end{table}

\begin{figure}
\centering
\includegraphics[width=\linewidth]{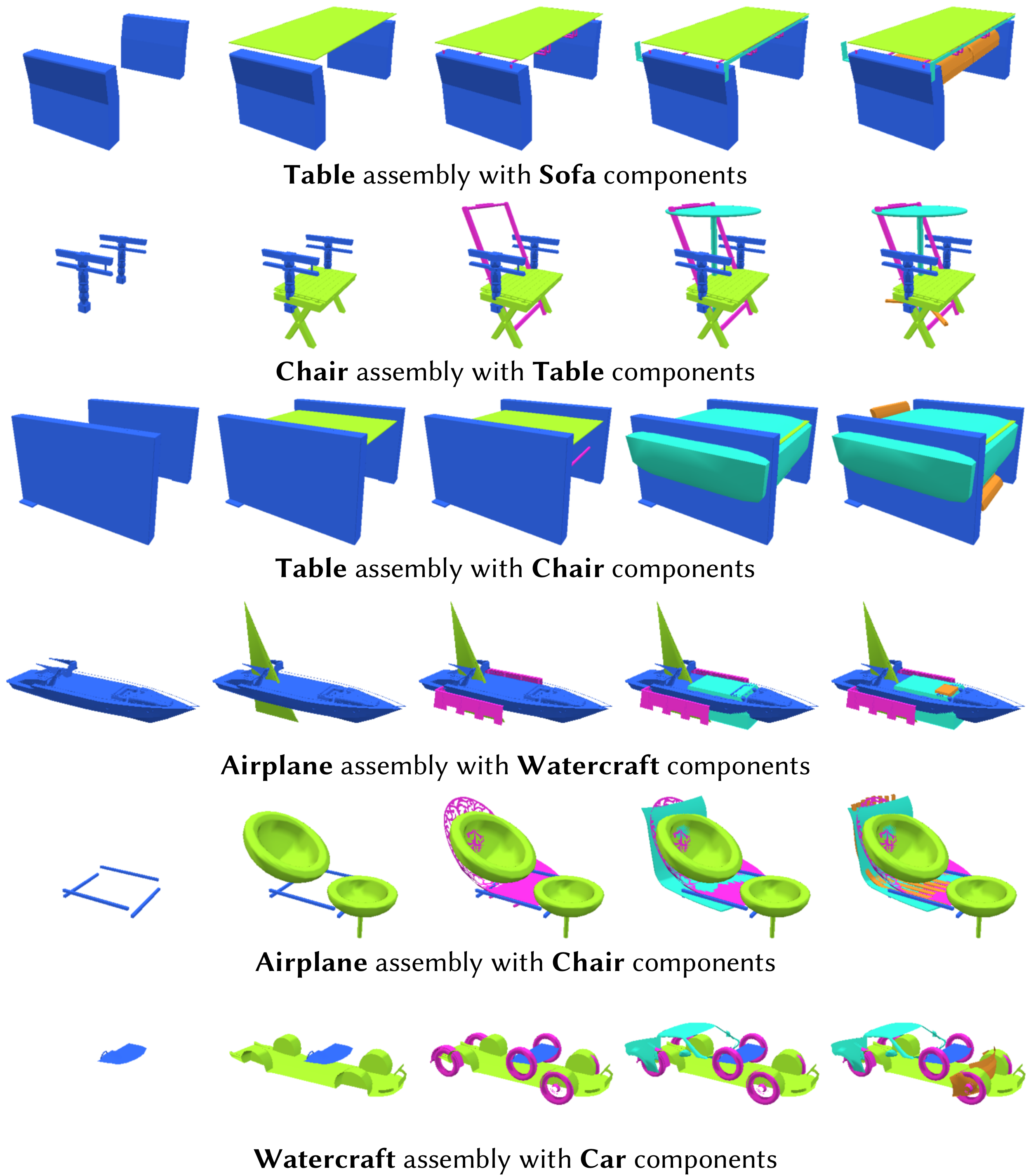}
\caption{\rev{Automatic assembly results using components in the other category. The assemblies across table, sofa, and chair categories are more reasonable than the assemblies across airplane, watercraft, and car categories due to the commonality in the component shapes.}}
\label{fig:cross_category_assembly}
\end{figure}

\vspace{0.1cm}
{\noindent \bfseries \rev{Cross-category assembly.}}
\rev{Lastly, we test to our method to assembly components in a different category with the trained models. \Fig{cross_category_assembly} demonstrates some results of cross-category automatic synthesis. We achieve reasonable outputs when using categories sharing many similar components (e.g., table - sofa - chair). Note that some components are even used with different functionalities such as table legs as chair arms and a back. Obviously, it is not possible to obtain plausible outputs when there is no commonality among component shapes (e.g., airplane - watercraft - car), but still the outputs show meaningful mappings such as a watercraft body to an airplane body and a watercraft sail to a airplane tail wing.}

\vspace{0.1cm}
{\noindent \bfseries Timing.}
We ran both training and test with a single NVIDIA GeForce GTX TITAN X Graphics Card. It took 12 hours to train each of the retrieval/embedding networks and placement networks for $100k$ iterations. In test time, each of the retrieval and placement network takes 0.1 second. 


\section{Conclusion}
\label{sec:conclusion}
We propose a novel method for retrieving and placing complementary parts for assembly-based modeling tools. Our method does not require consistent segmentation and part labeling, and it can learn from a non-curated and inconsistent oversegmentation of shapes in an online repository. We jointly learn how to predict complementary parts and how to organize them in a low-dimensional manifold. This enables us to retrieve parts that have good functional, geometric, and stylistic compatibility with the query shape. We also propose the first method to predict target part position just from their normalized geometry. 

\begin{figure}
\centering
\includegraphics[width=\linewidth]{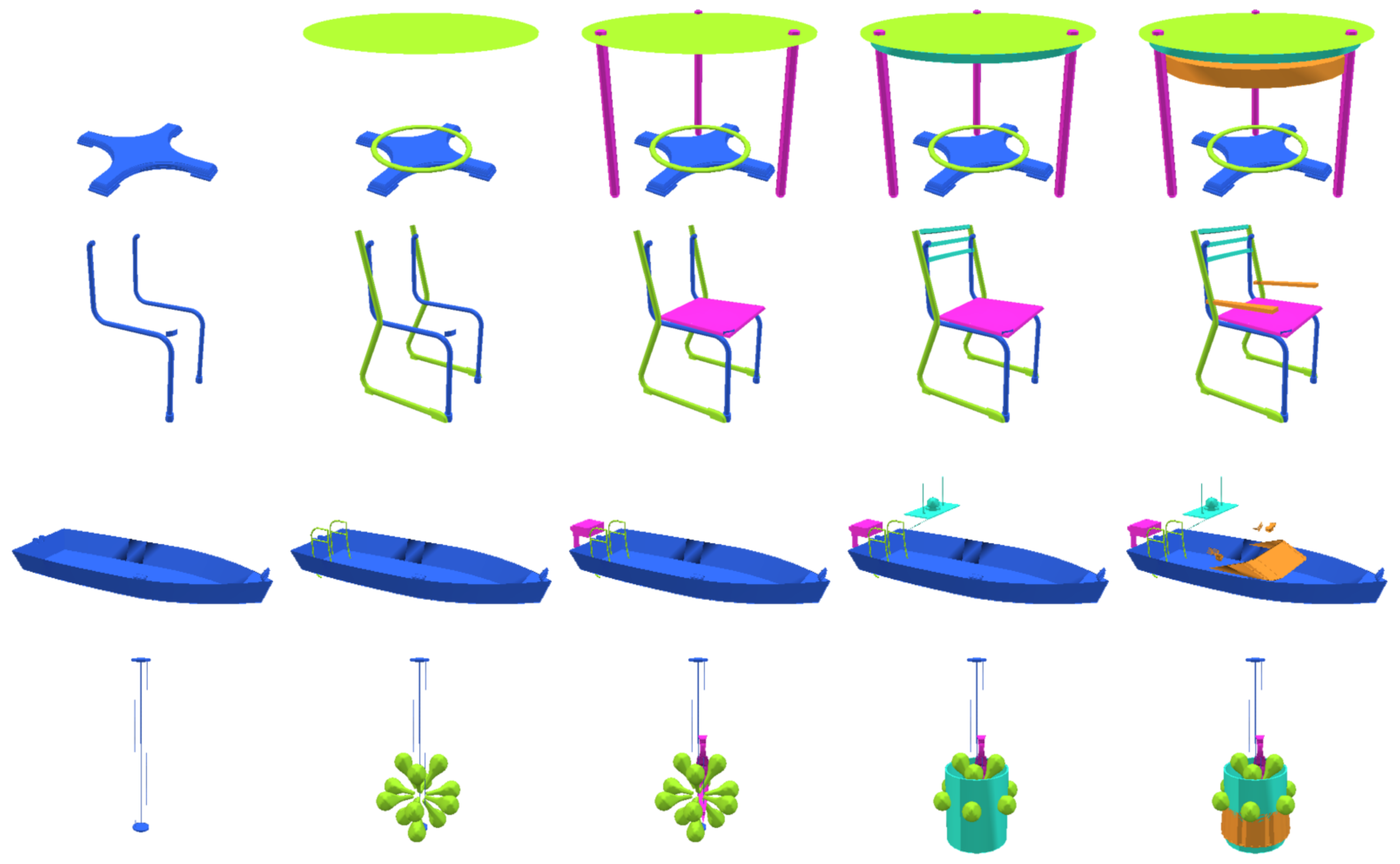}
\caption{\rev{Failure examples. 1st row: the bottom components match a single leg, but triple legs are retrieved. 2nd row: Green and blue components partially overlap. 3rd row: A sky-blue component is floating. 4th row: Components overlap.}}
\label{fig:failure_cases}
\end{figure}

\rev{Our framework has some limitations. While we randomly sample points over the surface of the partial input shape so that larger components have bigger influences in the next component retrieval, sometimes small/thin components play roles as certain parts, deciding the style of the whole object and occupying certain areas. Thus, the retrieval network can result in unreasonable outputs when these components are not well taken into account. In \Fig{failure_cases}, the automatically synthesized shapes have conflicted and unmatched components. Our placement network may also break physical constraints and have the new component to float or overlap with the input components. These issues, however, can be easily fixed with the user interaction.}

In the future, we plan to augment our method with capabilities to synthesize and deform retrieved parts, providing an even better compatibility with the query. For any practical interactive interface, it is essential to also provide additional user control beyond part selection: for example, enabling specifying high-level part attributes, rough shapes, and rough placements.


\bibliographystyle{ACM-Reference-Format}

\nocite{*}
\bibliography{paper}


\begin{thebibliography}{00}


\ifx \showCODEN    \undefined \def \showCODEN     #1{\unskip}     \fi
\ifx \showDOI      \undefined \def \showDOI       #1{{\tt DOI:}\penalty0{#1}\ }
  \fi
\ifx \showISBNx    \undefined \def \showISBNx     #1{\unskip}     \fi
\ifx \showISBNxiii \undefined \def \showISBNxiii  #1{\unskip}     \fi
\ifx \showISSN     \undefined \def \showISSN      #1{\unskip}     \fi
\ifx \showLCCN     \undefined \def \showLCCN      #1{\unskip}     \fi
\ifx \shownote     \undefined \def \shownote      #1{#1}          \fi
\ifx \showarticletitle \undefined \def \showarticletitle #1{#1}   \fi
\ifx \showURL      \undefined \def \showURL       {\relax}        \fi
\providecommand\bibfield[2]{#2}
\providecommand\bibinfo[2]{#2}
\providecommand\natexlab[1]{#1}
\providecommand\showeprint[2][]{arXiv:#2}

\bibitem[\protect\citeauthoryear{Bishop}{Bishop}{1994}]%
        {Bishop:1994}
\bibfield{author}{\bibinfo{person}{Christopher~M. Bishop}.}
  \bibinfo{year}{1994}\natexlab{}.
\newblock \bibinfo{booktitle}{{\em Mixture density networks}}.
\newblock \bibinfo{type}{{T}echnical {R}eport}. \bibinfo{institution}{Neural
  Computing Research Group, Aston University}.
\newblock


\bibitem[\protect\citeauthoryear{Chang, Funkhouser, Guibas, Hanrahan, Huang,
  Li, Savarese, Savva, Song, Su, Xiao, Yi, and Yu}{Chang et~al\mbox{.}}{2015}]%
        {Chang:2015}
\bibfield{author}{\bibinfo{person}{Angel~X. Chang}, \bibinfo{person}{Thomas~A.
  Funkhouser}, \bibinfo{person}{Leonidas~J. Guibas}, \bibinfo{person}{Pat
  Hanrahan}, \bibinfo{person}{Qi{-}Xing Huang}, \bibinfo{person}{Zimo Li},
  \bibinfo{person}{Silvio Savarese}, \bibinfo{person}{Manolis Savva},
  \bibinfo{person}{Shuran Song}, \bibinfo{person}{Hao Su},
  \bibinfo{person}{Jianxiong Xiao}, \bibinfo{person}{Li Yi}, {and}
  \bibinfo{person}{Fisher Yu}.} \bibinfo{year}{2015}\natexlab{}.
\newblock \showarticletitle{ShapeNet: An Information-Rich 3D Model Repository}.
\newblock \bibinfo{journal}{{\em CoRR\/}}  \bibinfo{volume}{abs/1512.03012}
  (\bibinfo{year}{2015}).
\newblock


\bibitem[\protect\citeauthoryear{Chaudhuri, Kalogerakis, Guibas, and
  Koltun}{Chaudhuri et~al\mbox{.}}{2011}]%
        {Chaudhuri:2011}
\bibfield{author}{\bibinfo{person}{Siddhartha Chaudhuri},
  \bibinfo{person}{Evangelos Kalogerakis}, \bibinfo{person}{Leonidas Guibas},
  {and} \bibinfo{person}{Vladlen Koltun}.} \bibinfo{year}{2011}\natexlab{}.
\newblock \showarticletitle{Probabilistic Reasoning for Assembly-based 3D
  Modeling}. In \bibinfo{booktitle}{{\em {SIGGRAPH}}}.
\newblock


\bibitem[\protect\citeauthoryear{Chaudhuri and Koltun}{Chaudhuri and
  Koltun}{2010}]%
        {Chaudhuri:2010}
\bibfield{author}{\bibinfo{person}{Siddhartha Chaudhuri} {and}
  \bibinfo{person}{Vladlen Koltun}.} \bibinfo{year}{2010}\natexlab{}.
\newblock \showarticletitle{Data-driven Suggestions for Creativity Support in
  3D Modeling}. In \bibinfo{booktitle}{{\em {SIGGRAPH Asia}}}.
\newblock


\bibitem[\protect\citeauthoryear{Chechik, Sharma, Shalit, and Bengio}{Chechik
  et~al\mbox{.}}{2010}]%
        {Chechik:2010}
\bibfield{author}{\bibinfo{person}{Gal Chechik}, \bibinfo{person}{Varun
  Sharma}, \bibinfo{person}{Uri Shalit}, {and} \bibinfo{person}{Samy Bengio}.}
  \bibinfo{year}{2010}\natexlab{}.
\newblock \showarticletitle{Large Scale Online Learning of Image Similarity
  Through Ranking}.
\newblock \bibinfo{journal}{{\em {JMLR}\/}} (\bibinfo{year}{2010}).
\newblock


\bibitem[\protect\citeauthoryear{Choy, Xu, Gwak, Chen, and Savarese}{Choy
  et~al\mbox{.}}{2016}]%
        {Choy:2016}
\bibfield{author}{\bibinfo{person}{Christopher~B. Choy},
  \bibinfo{person}{Danfei Xu}, \bibinfo{person}{JunYoung Gwak},
  \bibinfo{person}{Kevin Chen}, {and} \bibinfo{person}{Silvio Savarese}.}
  \bibinfo{year}{2016}\natexlab{}.
\newblock \bibinfo{booktitle}{{\em 3D-R2N2: A Unified Approach for Single and
  Multi-view 3D Object Reconstruction}}.
\newblock


\bibitem[\protect\citeauthoryear{Fan, Su, and Guibas}{Fan
  et~al\mbox{.}}{2017}]%
        {Fan:2017}
\bibfield{author}{\bibinfo{person}{Haoqiang Fan}, \bibinfo{person}{Hao Su},
  {and} \bibinfo{person}{Leonidas Guibas}.} \bibinfo{year}{2017}\natexlab{}.
\newblock \showarticletitle{A Point Set Generation Network for 3D Object
  Reconstruction from a Single Image}.
\newblock \bibinfo{journal}{{\em {CVPR}\/}} (\bibinfo{year}{2017}).
\newblock


\bibitem[\protect\citeauthoryear{Fisher, Ritchie, Savva, Funkhouser, and
  Hanrahan}{Fisher et~al\mbox{.}}{2012}]%
        {Fisher:2012}
\bibfield{author}{\bibinfo{person}{Matthew Fisher}, \bibinfo{person}{Daniel
  Ritchie}, \bibinfo{person}{Manolis Savva}, \bibinfo{person}{Thomas
  Funkhouser}, {and} \bibinfo{person}{Pat Hanrahan}.}
  \bibinfo{year}{2012}\natexlab{}.
\newblock \showarticletitle{Example-based Synthesis of 3D Object Arrangements}.
\newblock \bibinfo{journal}{{\em {SIGGRAPH Asia}\/}} (\bibinfo{year}{2012}).
\newblock


\bibitem[\protect\citeauthoryear{Funkhouser, Kazhdan, Shilane, Min, Kiefer,
  Tal, Rusinkiewicz, and Dobkin}{Funkhouser et~al\mbox{.}}{2004}]%
        {Funkhouser:2004}
\bibfield{author}{\bibinfo{person}{Thomas Funkhouser}, \bibinfo{person}{Michael
  Kazhdan}, \bibinfo{person}{Philip Shilane}, \bibinfo{person}{Patrick Min},
  \bibinfo{person}{William Kiefer}, \bibinfo{person}{Ayellet Tal},
  \bibinfo{person}{Szymon Rusinkiewicz}, {and} \bibinfo{person}{David Dobkin}.}
  \bibinfo{year}{2004}\natexlab{}.
\newblock \showarticletitle{Modeling by Example}. In \bibinfo{booktitle}{{\em
  {SIGGRAPH}}}.
\newblock


\bibitem[\protect\citeauthoryear{Garcia-Garcia, Gomez-Donoso, Garcia-Rodriguez,
  Orts-Escolano, Cazorla, and Azorin-Lopez}{Garcia-Garcia
  et~al\mbox{.}}{2016}]%
        {Garcia:2016}
\bibfield{author}{\bibinfo{person}{A. Garcia-Garcia}, \bibinfo{person}{F.
  Gomez-Donoso}, \bibinfo{person}{J. Garcia-Rodriguez}, \bibinfo{person}{S.
  Orts-Escolano}, \bibinfo{person}{M. Cazorla}, {and} \bibinfo{person}{J.
  Azorin-Lopez}.} \bibinfo{year}{2016}\natexlab{}.
\newblock \showarticletitle{PointNet: A 3D Convolutional Neural Network for
  real-time object class recognition}. In \bibinfo{booktitle}{{\em {IJCNN}}}.
\newblock


\bibitem[\protect\citeauthoryear{Girdhar, Fouhey, Rodriguez, and Gupta}{Girdhar
  et~al\mbox{.}}{2016}]%
        {Girdhar:2016}
\bibfield{author}{\bibinfo{person}{Rohit Girdhar}, \bibinfo{person}{David~F.
  Fouhey}, \bibinfo{person}{Mikel Rodriguez}, {and} \bibinfo{person}{Abhinav
  Gupta}.} \bibinfo{year}{2016}\natexlab{}.
\newblock \bibinfo{booktitle}{{\em Learning a Predictable and Generative Vector
  Representation for Objects}}.
\newblock


\bibitem[\protect\citeauthoryear{Golovinskiy and Funkhouser}{Golovinskiy and
  Funkhouser}{2008}]%
        {Golovinskiy:2008}
\bibfield{author}{\bibinfo{person}{Aleksey Golovinskiy} {and}
  \bibinfo{person}{Thomas Funkhouser}.} \bibinfo{year}{2008}\natexlab{}.
\newblock \showarticletitle{Randomized Cuts for {3D} Mesh Analysis}.
\newblock \bibinfo{journal}{{\em ACM Transactions on Graphics (Proc. SIGGRAPH
  ASIA)\/}} \bibinfo{volume}{27}, \bibinfo{number}{5} (\bibinfo{date}{Dec.}
  \bibinfo{year}{2008}).
\newblock


\bibitem[\protect\citeauthoryear{Golovinskiy and Funkhouser}{Golovinskiy and
  Funkhouser}{2009}]%
        {Golovinskiy:2009}
\bibfield{author}{\bibinfo{person}{Aleksey Golovinskiy} {and}
  \bibinfo{person}{Thomas Funkhouser}.} \bibinfo{year}{2009}\natexlab{}.
\newblock \showarticletitle{Consistent Segmentation of {3D} Models}.
\newblock \bibinfo{journal}{{\em {Proc. SMI}\/}} (\bibinfo{year}{2009}).
\newblock


\bibitem[\protect\citeauthoryear{Grant, Kohli, and van Gerven}{Grant
  et~al\mbox{.}}{2016}]%
        {Grant:2016}
\bibfield{author}{\bibinfo{person}{Edward Grant}, \bibinfo{person}{Pushmeet
  Kohli}, {and} \bibinfo{person}{Marcel van Gerven}.}
  \bibinfo{year}{2016}\natexlab{}.
\newblock \bibinfo{booktitle}{{\em Deep Disentangled Representations for
  Volumetric Reconstruction}}.
\newblock


\bibitem[\protect\citeauthoryear{Hadsell, Chopra, and LeCun}{Hadsell
  et~al\mbox{.}}{2006}]%
        {Hadsell:2006}
\bibfield{author}{\bibinfo{person}{R. Hadsell}, \bibinfo{person}{S. Chopra},
  {and} \bibinfo{person}{Y. LeCun}.} \bibinfo{year}{2006}\natexlab{}.
\newblock \showarticletitle{Dimensionality Reduction by Learning an Invariant
  Mapping}. In \bibinfo{booktitle}{{\em {CVPR}}}.
\newblock


\bibitem[\protect\citeauthoryear{Hu, Fan, , and Liu}{Hu et~al\mbox{.}}{2012}]%
        {Hu:2012}
\bibfield{author}{\bibinfo{person}{R. Hu}, \bibinfo{person}{L. Fan},
  \bibinfo{person}{}, {and} \bibinfo{person}{L. Liu}.}
  \bibinfo{year}{2012}\natexlab{}.
\newblock \showarticletitle{Co-segmentation of 3D shapes via subspace
  clustering}.
\newblock \bibinfo{journal}{{\em SGP\/}} (\bibinfo{year}{2012}).
\newblock


\bibitem[\protect\citeauthoryear{Hu, Zhu, van Kaick, Liu, Shamir, and Zhang}{Hu
  et~al\mbox{.}}{2015}]%
        {Hu:2015}
\bibfield{author}{\bibinfo{person}{Ruizhen Hu}, \bibinfo{person}{Chenyang Zhu},
  \bibinfo{person}{Oliver van Kaick}, \bibinfo{person}{Ligang Liu},
  \bibinfo{person}{Ariel Shamir}, {and} \bibinfo{person}{Hao Zhang}.}
  \bibinfo{year}{2015}\natexlab{}.
\newblock \showarticletitle{Interaction Context (ICON): Towards a Geometric
  Functionality Descriptor}.
\newblock \bibinfo{journal}{{\em {SIGGRAPH}\/}} (\bibinfo{year}{2015}).
\newblock


\bibitem[\protect\citeauthoryear{Huang, Koltun, and Guibas}{Huang
  et~al\mbox{.}}{2011}]%
        {Huang:2011}
\bibfield{author}{\bibinfo{person}{Qixing Huang}, \bibinfo{person}{Vladlen
  Koltun}, {and} \bibinfo{person}{Leonidas Guibas}.}
  \bibinfo{year}{2011}\natexlab{}.
\newblock \showarticletitle{Joint shape segmentation with linear programming}.
  In \bibinfo{booktitle}{{\em {SIGGRAPH Asia}}}.
\newblock


\bibitem[\protect\citeauthoryear{Huang, Wang, and Guibas}{Huang
  et~al\mbox{.}}{2014}]%
        {Huang:2014}
\bibfield{author}{\bibinfo{person}{Qixing Huang}, \bibinfo{person}{Fan Wang},
  {and} \bibinfo{person}{Leonidas Guibas}.} \bibinfo{year}{2014}\natexlab{}.
\newblock \showarticletitle{Functional Map Networks for Analyzing and Exploring
  Large Shape Collections}.
\newblock \bibinfo{journal}{{\em {ACM TOG}\/}}  \bibinfo{volume}{33}
  (\bibinfo{year}{2014}).
\newblock


\bibitem[\protect\citeauthoryear{Jaiswal, Huang, and Rai}{Jaiswal
  et~al\mbox{.}}{2016}]%
        {Jaiswal:2016}
\bibfield{author}{\bibinfo{person}{Prakhar Jaiswal}, \bibinfo{person}{Jinmiao
  Huang}, {and} \bibinfo{person}{Rahul Rai}.} \bibinfo{year}{2016}\natexlab{}.
\newblock \showarticletitle{Assembly-based conceptual 3D modeling with
  unlabeled components using probabilistic factor graph}.
\newblock \bibinfo{journal}{{\em Computer-Aided Design\/}}
  (\bibinfo{year}{2016}).
\newblock


\bibitem[\protect\citeauthoryear{Kalogerakis, Chaudhuri, Koller, and
  Koltun}{Kalogerakis et~al\mbox{.}}{2012}]%
        {Kalogerakis:2012}
\bibfield{author}{\bibinfo{person}{Evangelos Kalogerakis},
  \bibinfo{person}{Siddhartha Chaudhuri}, \bibinfo{person}{Daphne Koller},
  {and} \bibinfo{person}{Vladlen Koltun}.} \bibinfo{year}{2012}\natexlab{}.
\newblock \showarticletitle{A Probabilistic Model for Component-based Shape
  Synthesis}.
\newblock \bibinfo{journal}{{\em {SIGGRAPH}\/}} (\bibinfo{year}{2012}).
\newblock


\bibitem[\protect\citeauthoryear{Kalogerakis, Hertzmann, and Singh}{Kalogerakis
  et~al\mbox{.}}{2010}]%
        {Kalogerakis:2010}
\bibfield{author}{\bibinfo{person}{Evangelos Kalogerakis},
  \bibinfo{person}{Aaron Hertzmann}, {and} \bibinfo{person}{Karan Singh}.}
  \bibinfo{year}{2010}\natexlab{}.
\newblock \showarticletitle{Learning {3D} mesh segmentation and labeling}. In
  \bibinfo{booktitle}{{\em {SIGGRAPH}}}.
\newblock


\bibitem[\protect\citeauthoryear{Kim, Chaudhuri, Guibas, and Funkhouser}{Kim
  et~al\mbox{.}}{2014}]%
        {Kim:2014}
\bibfield{author}{\bibinfo{person}{Vladimir~G. Kim},
  \bibinfo{person}{Siddhartha Chaudhuri}, \bibinfo{person}{Leonidas Guibas},
  {and} \bibinfo{person}{Thomas Funkhouser}.} \bibinfo{year}{2014}\natexlab{}.
\newblock \showarticletitle{Shape2Pose: Human-centric Shape Analysis}.
\newblock \bibinfo{journal}{{\em {SIGGRAPH}\/}} (\bibinfo{year}{2014}).
\newblock


\bibitem[\protect\citeauthoryear{Kim, Li, Mitra, Chaudhuri, DiVerdi, and
  Funkhouser}{Kim et~al\mbox{.}}{2013}]%
        {Kim:2013a}
\bibfield{author}{\bibinfo{person}{Vladimir~G. Kim}, \bibinfo{person}{Wilmot
  Li}, \bibinfo{person}{Niloy~J. Mitra}, \bibinfo{person}{Siddhartha
  Chaudhuri}, \bibinfo{person}{Stephen DiVerdi}, {and} \bibinfo{person}{Thomas
  Funkhouser}.} \bibinfo{year}{2013}\natexlab{}.
\newblock \showarticletitle{Learning Part-based Templates from Large
  Collections of 3D Shapes}.
\newblock \bibinfo{journal}{{\em {SIGGRAPH}\/}} (\bibinfo{year}{2013}).
\newblock


\bibitem[\protect\citeauthoryear{Kingma and Ba}{Kingma and Ba}{2014}]%
        {Kingma:2014b}
\bibfield{author}{\bibinfo{person}{Diederik~P. Kingma} {and}
  \bibinfo{person}{Jimmy Ba}.} \bibinfo{year}{2014}\natexlab{}.
\newblock \showarticletitle{Adam: {A} Method for Stochastic Optimization}.
\newblock \bibinfo{journal}{{\em CoRR\/}} (\bibinfo{year}{2014}).
\newblock


\bibitem[\protect\citeauthoryear{Kingma, Rezende, Mohamed, and Welling}{Kingma
  et~al\mbox{.}}{2014}]%
        {Kingma:2014}
\bibfield{author}{\bibinfo{person}{Diederik~P. Kingma},
  \bibinfo{person}{Danilo~J. Rezende}, \bibinfo{person}{Shakir Mohamed}, {and}
  \bibinfo{person}{Max Welling}.} \bibinfo{year}{2014}\natexlab{}.
\newblock \showarticletitle{Semi-supervised Learning with Deep Generative
  Models}. In \bibinfo{booktitle}{{\em {NIPS}}}.
\newblock


\bibitem[\protect\citeauthoryear{Kong, Lin, , and Lucey}{Kong
  et~al\mbox{.}}{2017}]%
        {Kong:2017}
\bibfield{author}{\bibinfo{person}{Chen Kong}, \bibinfo{person}{Chen-Hsuan
  Lin}, \bibinfo{person}{}, {and} \bibinfo{person}{Simon Lucey}.}
  \bibinfo{year}{2017}\natexlab{}.
\newblock \showarticletitle{Using Locally Corresponding CAD Models for Dense 3D
  Reconstructions from a Single Image}.
\newblock \bibinfo{journal}{{\em {CVPR}\/}} (\bibinfo{year}{2017}).
\newblock


\bibitem[\protect\citeauthoryear{Li, Xu, Chaudhuri, Yumer, Zhang, and
  Guibas}{Li et~al\mbox{.}}{2017}]%
        {Li:2017}
\bibfield{author}{\bibinfo{person}{Jun Li}, \bibinfo{person}{Kai Xu},
  \bibinfo{person}{Siddhartha Chaudhuri}, \bibinfo{person}{Ersin Yumer},
  \bibinfo{person}{Hao Zhang}, {and} \bibinfo{person}{Leonidas Guibas}.}
  \bibinfo{year}{2017}\natexlab{}.
\newblock \showarticletitle{GRASS: Generative Recursive Autoencoders for Shape
  Structures}.
\newblock \bibinfo{journal}{{\em {SIGGRAPH}\/}} (\bibinfo{year}{2017}).
\newblock


\bibitem[\protect\citeauthoryear{Li, Dai, Guibas, and Nie{\ss}ner}{Li
  et~al\mbox{.}}{2015}]%
        {Li:2015}
\bibfield{author}{\bibinfo{person}{Yangyan Li}, \bibinfo{person}{Angela Dai},
  \bibinfo{person}{Leonidas Guibas}, {and} \bibinfo{person}{Matthias
  Nie{\ss}ner}.} \bibinfo{year}{2015}\natexlab{}.
\newblock \showarticletitle{Database-Assisted Object Retrieval for Real-Time 3D
  Reconstruction}.
\newblock  (\bibinfo{year}{2015}).
\newblock


\bibitem[\protect\citeauthoryear{Makadia and Yumer}{Makadia and Yumer}{2014}]%
        {Makadia:2014}
\bibfield{author}{\bibinfo{person}{A. Makadia} {and} \bibinfo{person}{M.~E.
  Yumer}.} \bibinfo{year}{2014}\natexlab{}.
\newblock \showarticletitle{Learning 3D Part Detection from Sparsely Labeled
  Data}. In \bibinfo{booktitle}{{\em 3DV}}.
\newblock


\bibitem[\protect\citeauthoryear{Mirza and Osindero}{Mirza and
  Osindero}{2014}]%
        {Mirza:2014}
\bibfield{author}{\bibinfo{person}{Mehdi Mirza} {and} \bibinfo{person}{Simon
  Osindero}.} \bibinfo{year}{2014}\natexlab{}.
\newblock \showarticletitle{Conditional Generative Adversarial Nets}.
\newblock \bibinfo{journal}{{\em CoRR\/}}  \bibinfo{volume}{abs/1411.1784}
  (\bibinfo{year}{2014}).
\newblock


\bibitem[\protect\citeauthoryear{Osada, Funkhouser, Chazelle, and Dobkin}{Osada
  et~al\mbox{.}}{2002}]%
        {Osada:2002}
\bibfield{author}{\bibinfo{person}{Robert Osada}, \bibinfo{person}{Thomas
  Funkhouser}, \bibinfo{person}{Bernard Chazelle}, {and} \bibinfo{person}{David
  Dobkin}.} \bibinfo{year}{2002}\natexlab{}.
\newblock \showarticletitle{Shape Distributions}.
\newblock \bibinfo{journal}{{\em {ACM TOG}\/}} (\bibinfo{year}{2002}).
\newblock


\bibitem[\protect\citeauthoryear{Qi, Su, Mo, and Guibas}{Qi
  et~al\mbox{.}}{2017}]%
        {Qi:2017}
\bibfield{author}{\bibinfo{person}{Charles~Ruizhongtai Qi},
  \bibinfo{person}{Hao Su}, \bibinfo{person}{Kaichun Mo}, {and}
  \bibinfo{person}{Leonidas~J. Guibas}.} \bibinfo{year}{2017}\natexlab{}.
\newblock \showarticletitle{PointNet: Deep Learning on Point Sets for 3D
  Classification and Segmentation}.
\newblock \bibinfo{journal}{{\em {CVPR}\/}} (\bibinfo{year}{2017}).
\newblock


\bibitem[\protect\citeauthoryear{Qi, Su, Nießner, Dai, Yan, and Guibas}{Qi
  et~al\mbox{.}}{2016}]%
        {Qi:2016}
\bibfield{author}{\bibinfo{person}{C.~R. Qi}, \bibinfo{person}{H. Su},
  \bibinfo{person}{M. Nießner}, \bibinfo{person}{A. Dai}, \bibinfo{person}{M.
  Yan}, {and} \bibinfo{person}{L.~J. Guibas}.} \bibinfo{year}{2016}\natexlab{}.
\newblock \showarticletitle{Volumetric and Multi-view CNNs for Object
  Classification on 3D Data}. In \bibinfo{booktitle}{{\em {CVPR}}}.
\newblock


\bibitem[\protect\citeauthoryear{Riegler, Ulusoy, and Geiger}{Riegler
  et~al\mbox{.}}{2017}]%
        {Riegler:2017}
\bibfield{author}{\bibinfo{person}{Gernot Riegler}, \bibinfo{person}{Ali~Osman
  Ulusoy}, {and} \bibinfo{person}{Andreas Geiger}.}
  \bibinfo{year}{2017}\natexlab{}.
\newblock \showarticletitle{OctNet: Learning Deep 3D Representations at High
  Resolutions}.
\newblock \bibinfo{journal}{{\em {CVPR}\/}} (\bibinfo{year}{2017}).
\newblock


\bibitem[\protect\citeauthoryear{Shao, Monszpart, Zheng, Koo, Xu, Zhou, and
  Mitra}{Shao et~al\mbox{.}}{2014}]%
        {Shao:2014}
\bibfield{author}{\bibinfo{person}{Tianjia Shao}, \bibinfo{person}{Aron
  Monszpart}, \bibinfo{person}{Youyi Zheng}, \bibinfo{person}{Bongjin Koo},
  \bibinfo{person}{Weiwei Xu}, \bibinfo{person}{Kun Zhou}, {and}
  \bibinfo{person}{Niloy~J. Mitra}.} \bibinfo{year}{2014}\natexlab{}.
\newblock \showarticletitle{Imagining the Unseen: Stability-based Cuboid
  Arrangements for Scene Understanding}.
\newblock \bibinfo{journal}{{\em {SIGGRAPH Asia}\/}} (\bibinfo{year}{2014}).
\newblock


\bibitem[\protect\citeauthoryear{Sharma, Grau, and Fritz}{Sharma
  et~al\mbox{.}}{2016}]%
        {Sharma:2016}
\bibfield{author}{\bibinfo{person}{Abhishek Sharma}, \bibinfo{person}{Oliver
  Grau}, {and} \bibinfo{person}{Mario Fritz}.} \bibinfo{year}{2016}\natexlab{}.
\newblock \bibinfo{booktitle}{{\em VConv-DAE: Deep Volumetric Shape Learning
  Without Object Labels}}.
\newblock


\bibitem[\protect\citeauthoryear{Shen, Fu, Chen, and Hu}{Shen
  et~al\mbox{.}}{2012}]%
        {Shen:2012}
\bibfield{author}{\bibinfo{person}{Chao-Hui Shen}, \bibinfo{person}{Hongbo Fu},
  \bibinfo{person}{Kang Chen}, {and} \bibinfo{person}{Shi-Min Hu}.}
  \bibinfo{year}{2012}\natexlab{}.
\newblock \showarticletitle{Structure Recovery by Part Assembly}.
\newblock \bibinfo{journal}{{\em {SIGGRAPH Asia}\/}} (\bibinfo{year}{2012}).
\newblock


\bibitem[\protect\citeauthoryear{Sidi, van Kaick, Kleiman, Zhang, and
  Cohen-Or}{Sidi et~al\mbox{.}}{2011}]%
        {Sidi:2011}
\bibfield{author}{\bibinfo{person}{Oana Sidi}, \bibinfo{person}{Oliver van
  Kaick}, \bibinfo{person}{Yanir Kleiman}, \bibinfo{person}{Hao Zhang}, {and}
  \bibinfo{person}{Daniel Cohen-Or}.} \bibinfo{year}{2011}\natexlab{}.
\newblock \showarticletitle{Unsupervised Co-Segmentation of a Set of Shapes via
  Descriptor-Space Spectral Clustering}.
\newblock \bibinfo{journal}{{\em {SIGGRAPH Asia}\/}} (\bibinfo{year}{2011}).
\newblock


\bibitem[\protect\citeauthoryear{Sohn, Yan, and Lee}{Sohn
  et~al\mbox{.}}{2015}]%
        {Sohn:2015}
\bibfield{author}{\bibinfo{person}{Kihyuk Sohn}, \bibinfo{person}{Xinchen Yan},
  {and} \bibinfo{person}{Honglak Lee}.} \bibinfo{year}{2015}\natexlab{}.
\newblock \showarticletitle{Learning Structured Output Representation Using
  Deep Conditional Generative Models}. In \bibinfo{booktitle}{{\em {NIPS}}}.
\newblock


\bibitem[\protect\citeauthoryear{Su, Li, Qi, Fish, Cohen-Or, and Guibas}{Su
  et~al\mbox{.}}{2015a}]%
        {su2015joint}
\bibfield{author}{\bibinfo{person}{Hao Su}, \bibinfo{person}{Yangyan Li},
  \bibinfo{person}{Charles~Ruizhongtai Qi}, \bibinfo{person}{Noa Fish},
  \bibinfo{person}{Daniel Cohen-Or}, {and} \bibinfo{person}{Leonidas~J
  Guibas}.} \bibinfo{year}{2015}\natexlab{a}.
\newblock \showarticletitle{Joint embeddings of shapes and images via cnn image
  purification}.
\newblock \bibinfo{journal}{{\em ACM Transactions on Graphics (TOG)\/}}
  \bibinfo{volume}{34}, \bibinfo{number}{6} (\bibinfo{year}{2015}),
  \bibinfo{pages}{234}.
\newblock


\bibitem[\protect\citeauthoryear{Su, Maji, Kalogerakis, and Learned-Miller}{Su
  et~al\mbox{.}}{2015b}]%
        {Su:2015}
\bibfield{author}{\bibinfo{person}{Hang Su}, \bibinfo{person}{Subhransu Maji},
  \bibinfo{person}{Evangelos Kalogerakis}, {and} \bibinfo{person}{Erik
  Learned-Miller}.} \bibinfo{year}{2015}\natexlab{b}.
\newblock \showarticletitle{Multi-view Convolutional Neural Networks for 3D
  Shape Recognition}. In \bibinfo{booktitle}{{\em {ICCV}}}.
\newblock


\bibitem[\protect\citeauthoryear{Sung, Kim, Angst, and Guibas}{Sung
  et~al\mbox{.}}{2015}]%
        {Sung:2015}
\bibfield{author}{\bibinfo{person}{Minhyuk Sung}, \bibinfo{person}{Vladimir~G.
  Kim}, \bibinfo{person}{Roland Angst}, {and} \bibinfo{person}{Leonidas
  Guibas}.} \bibinfo{year}{2015}\natexlab{}.
\newblock \showarticletitle{Data-driven Structural Priors for Shape
  Completion}.
\newblock \bibinfo{journal}{{\em {SIGGRAPH Asia}\/}} (\bibinfo{year}{2015}).
\newblock


\bibitem[\protect\citeauthoryear{Trimble}{Trimble}{2017}]%
        {Warehouse}
\bibfield{author}{\bibinfo{person}{Trimble}.} \bibinfo{year}{2017}\natexlab{}.
\newblock \showarticletitle{3D Warehouse}.
\newblock  (\bibinfo{year}{2017}).
\newblock
\showURL{%
\url{https://3dwarehouse.sketchup.com/}}


\bibitem[\protect\citeauthoryear{Tulsiani, Zhou, Efros, and Malik}{Tulsiani
  et~al\mbox{.}}{2017}]%
        {Tulsiani:2017}
\bibfield{author}{\bibinfo{person}{Shubham Tulsiani}, \bibinfo{person}{Tinghui
  Zhou}, \bibinfo{person}{Alexei~A. Efros}, {and} \bibinfo{person}{Jitendra
  Malik}.} \bibinfo{year}{2017}\natexlab{}.
\newblock \showarticletitle{Multi-view Supervision for Single-view
  Reconstruction via Differentiable Ray Consistency}. In
  \bibinfo{booktitle}{{\em {CVPR}}}.
\newblock


\bibitem[\protect\citeauthoryear{van~der Maaten}{van~der Maaten}{2009}]%
        {Maaten:2009}
\bibfield{author}{\bibinfo{person}{Laurens van~der Maaten}.}
  \bibinfo{year}{2009}\natexlab{}.
\newblock \showarticletitle{Learning a Parametric Embedding by Preserving Local
  Structure}. In \bibinfo{booktitle}{{\em {AISTATS}}}.
\newblock


\bibitem[\protect\citeauthoryear{van Kaick, Xu, Zhang, Wang, Sun, Shamir, and
  Cohen-Or}{van Kaick et~al\mbox{.}}{2013}]%
        {Kaick:2013}
\bibfield{author}{\bibinfo{person}{Oliver van Kaick}, \bibinfo{person}{Kai Xu},
  \bibinfo{person}{Hao Zhang}, \bibinfo{person}{Yanzhen Wang},
  \bibinfo{person}{Shuyang Sun}, \bibinfo{person}{Ariel Shamir}, {and}
  \bibinfo{person}{Daniel Cohen-Or}.} \bibinfo{year}{2013}\natexlab{}.
\newblock \showarticletitle{Co-hierarchical analysis of shape structures}.
\newblock \bibinfo{journal}{{\em {ACM TOG}\/}} (\bibinfo{year}{2013}).
\newblock


\bibitem[\protect\citeauthoryear{Wang, Asafi, van Kaick, Zhang, Cohen-Or, and
  Chen}{Wang et~al\mbox{.}}{2012}]%
        {Wang:2012}
\bibfield{author}{\bibinfo{person}{Yunhai Wang}, \bibinfo{person}{Shmulik
  Asafi}, \bibinfo{person}{Oliver van Kaick}, \bibinfo{person}{Hao Zhang},
  \bibinfo{person}{Daniel Cohen-Or}, {and} \bibinfo{person}{Baoquan Chen}.}
  \bibinfo{year}{2012}\natexlab{}.
\newblock \showarticletitle{Active Co-analysis of a Set of Shapes}.
\newblock \bibinfo{journal}{{\em {SIGGRAPH Asia}\/}} (\bibinfo{year}{2012}).
\newblock


\bibitem[\protect\citeauthoryear{Wu, Xue, Lim, Tian, Tenenbaum, Torralba, and
  Freeman}{Wu et~al\mbox{.}}{2016a}]%
        {Wu:2016c}
\bibfield{author}{\bibinfo{person}{Jiajun Wu}, \bibinfo{person}{Tianfan Xue},
  \bibinfo{person}{Joseph~J. Lim}, \bibinfo{person}{Yuandong Tian},
  \bibinfo{person}{Joshua~B. Tenenbaum}, \bibinfo{person}{Antonio Torralba},
  {and} \bibinfo{person}{William~T. Freeman}.}
  \bibinfo{year}{2016}\natexlab{a}.
\newblock \bibinfo{booktitle}{{\em Single Image 3D Interpreter Network}}.
\newblock


\bibitem[\protect\citeauthoryear{Wu, Zhang, Xue, Freeman, and Tenenbaum}{Wu
  et~al\mbox{.}}{2016b}]%
        {Wu:2016b}
\bibfield{author}{\bibinfo{person}{Jiajun Wu}, \bibinfo{person}{Chengkai
  Zhang}, \bibinfo{person}{Tianfan Xue}, \bibinfo{person}{William~T Freeman},
  {and} \bibinfo{person}{Joshua~B Tenenbaum}.}
  \bibinfo{year}{2016}\natexlab{b}.
\newblock \showarticletitle{Learning a probabilistic latent space of object
  shapes via 3d generative-adversarial modeling}. In \bibinfo{booktitle}{{\em
  {NIPS}}}.
\newblock


\bibitem[\protect\citeauthoryear{Wu, Song, Khosla, Tang, and Xiao}{Wu
  et~al\mbox{.}}{2015a}]%
        {Wu:2015}
\bibfield{author}{\bibinfo{person}{Zhirong Wu}, \bibinfo{person}{Shuran Song},
  \bibinfo{person}{Aditya Khosla}, \bibinfo{person}{Xiaoou Tang}, {and}
  \bibinfo{person}{Jianxiong Xiao}.} \bibinfo{year}{2015}\natexlab{a}.
\newblock \showarticletitle{3D ShapeNets for 2.5D Object Recognition and
  Next-Best-View Prediction}.
\newblock \bibinfo{journal}{{\em {CVPR}\/}} (\bibinfo{year}{2015}).
\newblock


\bibitem[\protect\citeauthoryear{Wu, Song, Khosla, Yu, Zhang, Tang, and
  Xiao}{Wu et~al\mbox{.}}{2015b}]%
        {Wu:2016}
\bibfield{author}{\bibinfo{person}{Zhirong Wu}, \bibinfo{person}{S. Song},
  \bibinfo{person}{A. Khosla}, \bibinfo{person}{Fisher Yu},
  \bibinfo{person}{Linguang Zhang}, \bibinfo{person}{Xiaoou Tang}, {and}
  \bibinfo{person}{J. Xiao}.} \bibinfo{year}{2015}\natexlab{b}.
\newblock \showarticletitle{3D ShapeNets: A deep representation for volumetric
  shapes}. In \bibinfo{booktitle}{{\em {CVPR}}}.
\newblock


\bibitem[\protect\citeauthoryear{Xie, Xu, Liu, and Xiong}{Xie
  et~al\mbox{.}}{2014}]%
        {Xie:2014}
\bibfield{author}{\bibinfo{person}{Zhige Xie}, \bibinfo{person}{Kai Xu},
  \bibinfo{person}{Ligang Liu}, {and} \bibinfo{person}{Yueshan Xiong}.}
  \bibinfo{year}{2014}\natexlab{}.
\newblock \showarticletitle{3D Shape Segmentation and Labeling via Extreme
  Learning Machine}.
\newblock \bibinfo{journal}{{\em SGP\/}} (\bibinfo{year}{2014}).
\newblock


\bibitem[\protect\citeauthoryear{Xu, Zhang, Cohen-Or, and Chen}{Xu
  et~al\mbox{.}}{2012}]%
        {Xu:2012}
\bibfield{author}{\bibinfo{person}{Kai Xu}, \bibinfo{person}{Hao Zhang},
  \bibinfo{person}{Daniel Cohen-Or}, {and} \bibinfo{person}{Baoquan Chen}.}
  \bibinfo{year}{2012}\natexlab{}.
\newblock \showarticletitle{Fit and Diverse: Set Evolution for Inspiring 3D
  Shape Galleries}.
\newblock \bibinfo{journal}{{\em {SIGGRAPH}\/}} (\bibinfo{year}{2012}).
\newblock


\bibitem[\protect\citeauthoryear{Yi, Guibas, Hertzmann, Kim, Su, and Yumer}{Yi
  et~al\mbox{.}}{2017}]%
        {Yi:2017}
\bibfield{author}{\bibinfo{person}{Li Yi}, \bibinfo{person}{Leonidas Guibas},
  \bibinfo{person}{Aaron Hertzmann}, \bibinfo{person}{Vladimir~G. Kim},
  \bibinfo{person}{Hao Su}, {and} \bibinfo{person}{Ersin Yumer}.}
  \bibinfo{year}{2017}\natexlab{}.
\newblock \showarticletitle{Learning Hierarchical Shape Segmentation and
  Labeling from Online Repositories}.
\newblock \bibinfo{journal}{{\em {SIGGRAPH}\/}} (\bibinfo{year}{2017}).
\newblock


\bibitem[\protect\citeauthoryear{Yi, Kim, Ceylan, Shen, Yan, Su, Lu, Huang,
  Sheffer, and Guibas}{Yi et~al\mbox{.}}{2016}]%
        {yi:2016}
\bibfield{author}{\bibinfo{person}{Li Yi}, \bibinfo{person}{Vladimir~G. Kim},
  \bibinfo{person}{Duygu Ceylan}, \bibinfo{person}{I-Chao Shen},
  \bibinfo{person}{Mengyan Yan}, \bibinfo{person}{Hao Su},
  \bibinfo{person}{Cewu Lu}, \bibinfo{person}{Qixing Huang},
  \bibinfo{person}{Alla Sheffer}, {and} \bibinfo{person}{Leonidas Guibas}.}
  \bibinfo{year}{2016}\natexlab{}.
\newblock \showarticletitle{A Scalable Active Framework for Region Annotation
  in 3D Shape Collections}.
\newblock \bibinfo{journal}{{\em {SIGGRAPH Asia}\/}} (\bibinfo{year}{2016}).
\newblock


\bibitem[\protect\citeauthoryear{Zheng, Cohen-Or, Averkiou, and Mitra}{Zheng
  et~al\mbox{.}}{2014}]%
        {Zheng:2014}
\bibfield{author}{\bibinfo{person}{Youyi Zheng}, \bibinfo{person}{Daniel
  Cohen-Or}, \bibinfo{person}{Melinos Averkiou}, {and}
  \bibinfo{person}{Niloy~J. Mitra}.} \bibinfo{year}{2014}\natexlab{}.
\newblock \showarticletitle{Recurring part arrangements in shape collections}.
\newblock  (\bibinfo{year}{2014}).
\newblock


\bibitem[\protect\citeauthoryear{Zheng, Cohen-Or, and Mitra}{Zheng
  et~al\mbox{.}}{2013}]%
        {Zheng:2013}
\bibfield{author}{\bibinfo{person}{Youyi Zheng}, \bibinfo{person}{Daniel
  Cohen-Or}, {and} \bibinfo{person}{Niloy~J. Mitra}.}
  \bibinfo{year}{2013}\natexlab{}.
\newblock \showarticletitle{Smart Variations: Functional Substructures for Part
  Compatibility}.
\newblock \bibinfo{journal}{{\em {Eurographics}\/}} (\bibinfo{year}{2013}).
\newblock


\end{thebibliography}
\appendix
\section{Appendix}
\label{sec:appendix}

\Eq{loss_joint} is expanded as follows:

\begin{alignat*}{2}
&E(X, Y) \\
&= -\log \sum_{k=1}^K \phi_k(X) && \mathcal{N}(f(Y) \,|\, \mu_k(X), \sigma_k(X)^2) \\
&= - \log \sum_{k=1}^K \phi_k(X) && \prod_{d=1}^{D} \left( \frac{1}{\sqrt{2 \pi} \sigma_{k,d}(X)} \exp \left( -\frac{ (f(Y)_{d} - \mu_{k,d}(X))^2 }{2 \sigma_{k,d}(X)^2} \right) \right) \\
&= - \log \sum_{k=1}^K \exp \Bigg[ && -\sum_{d=1}^{D} \left( \frac{ (f(Y)_{d} - \mu_{k,d}(X))^2 }{2 \sigma_{k,d}(X)^2} - \log \sigma_{k,d}(X) \right) \\
& && - D \frac{\log 2 \pi}{2} + \log \phi_k(X) \Bigg],
\end{alignat*}

where $x_d$ is the d-th dimension value of $x$. This leads to a \emph{log-sum-exp} formulation, which requires the following computation trick for avoiding numerical issues: 

\begin{align*}
\log \sum_k \exp \left[ x_k \right] = \max_k (x_k) + \log \sum_k \exp \left[ x_k - \max_k (x_k) \right]
\end{align*}

\end{document}